\documentclass{pasa}%

\usepackage{amsfonts}
\usepackage{amssymb}
\usepackage{amsmath}

\usepackage{aas_macros}

\usepackage{graphicx}

\usepackage[colorlinks]{hyperref}
\usepackage{etoolbox}

\hypersetup{colorlinks=true,linkcolor=blue,citecolor=blue,filecolor=blue,urlcolor=blue}
\makeatletter
\patchcmd{\NAT@citex}
  {\@citea\NAT@hyper@{%
     \NAT@nmfmt{\NAT@nm}%
     \hyper@natlinkbreak{\NAT@aysep\NAT@spacechar}{\@citeb\@extra@b@citeb}%
     \NAT@date}}
  {\@citea\NAT@nmfmt{\NAT@nm}%
   \NAT@aysep\NAT@spacechar\NAT@hyper@{\NAT@date}}{}{}

\patchcmd{\NAT@citex}
  {\@citea\NAT@hyper@{%
     \NAT@nmfmt{\NAT@nm}%
     \hyper@natlinkbreak{\NAT@spacechar\NAT@@open\if*#1*\else#1\NAT@spacechar\fi}%
       {\@citeb\@extra@b@citeb}%
     \NAT@date}}
  {\@citea\NAT@nmfmt{\NAT@nm}%
   \NAT@spacechar\NAT@@open\if*#1*\else#1\NAT@spacechar\fi\NAT@hyper@{\NAT@date}}
  {}{}
  
\def\aj{AJ}

\def\apj{ApJ}
\def\apjl{ApJ}
\def\apjs{ApJS}
\def\ao{Appl.~Opt.}
\def\apss{Ap\&SS}
\def\aap{A\&A}

\def\aaps{A\&AS}

\def\mnras{MNRAS}

\def\pasa{Publ.~Astron.~Soc.~Australia}
\def\pasp{PASP}

\newcommand{\eld}{${\it N}_{\rm e}$}
\newcommand{\hyd}{${\it N}{\rm (H^{+})}$}

\newcommand{\elt}{${\it T}_{\rm e}$}

\newcommand{\teff}{$T_{\rm eff}$}

\newcommand{\cmt}{cm$^{-3}$}

\newcommand{\foiii}{[O~{\sc iii}]}

\newcommand{\foi}{[O~{\sc i}]}
\newcommand{\foii}{[O~{\sc ii}]}

\newcommand{\fsii}{[S~{\sc ii}]}
\newcommand{\fsiii}{[S~{\sc iii}]}

\newcommand{\fni}{[N~{\sc i}]}
\newcommand{\fnii}{[N~{\sc ii}]}

\newcommand{\fariii}{[Ar~{\sc iii}]}

\newcommand{\fcliii}{[Cl~{\sc iii}]}

\newcommand{\fneii}{[Ne~{\sc ii}]}
\newcommand{\fneiii}{[Ne~{\sc iii}]}

\newcommand{\nii}{N~{\sc ii}}

\newcommand{\oii}{O~{\sc ii}}
\newcommand{\cii}{C~{\sc ii}}
\newcommand{\neii}{Ne~{\sc ii}}

\newcommand{\hi}{H\,{\sc i}}

\newcommand{\hei}{He~{\sc i}}

\newcommand{\ha}{H$\alpha$}
\newcommand{\hb}{H$\beta$}
\newcommand{\hg}{H$\gamma$}
\newcommand{\hd}{H$\delta$}

\def\p0{\phantom{0}}

\def\tabliterature1abund{14}

\title[Bi-abundance Ionization of PB~8]{Bi-Abundance Ionization Structure of the Wolf-Rayet Planetary Nebula PB~8}
\author[A.~Danehkar]{A.~Danehkar$^{1,2,3}$\\
\affil{$^1$Department of Physics and Astronomy, Macquarie University, Sydney, NSW 2109, Australia}%
\affil{$^2$Harvard-Smithsonian Center for Astrophysics, 60 Garden Street, Cambridge, MA 02138, USA}%
\affil{$^3$Email: \href{mailto:ashkbiz.danehkar@cfa.harvard.edu}{ashkbiz.danehkar@cfa.harvard.edu}}%
~\newline
{\footnotesize({\sc Received} August 29, 2017; {\sc Accepted} January 3, 2018)}
}%

\jid{PASA}
\doi{10.1017/pas.\the\year.xxx}
\jyear{\the\year}

\begin{document}%

\begin{frontmatter}
\maketitle

\begin{abstract}
The planetary nebula (PN) PB 8 around a [WN/WC]-hybrid central star is one of PNe with moderate abundance discrepancy factors (ADFs\,$\sim$\,2--3), which could be an indication of a tiny fraction of metal-rich inclusions embedded in the nebula (bi-abundance). In this work, we have constructed photoionization models to reproduce the optical and infrared observations of the PN PB 8 using a non-LTE stellar model atmosphere ionizing source. A chemically homogeneous model initially used cannot predict the optical recombination lines (ORLs). However, a bi-abundance model provides a better fit to most of the observed ORLs from N and O ions. The metal-rich inclusions in the bi-abundance model occupy 5.6 percent of the total volume of the nebula, and are roughly 1.7 times cooler and denser than the mean values of the surrounding nebula. The N/H and O/H  abundance ratios in the metal-rich inclusions are $\sim$ 1.0 and 1.7 dex larger than the diffuse warm nebula, respectively. To reproduce the \textit{Spitzer} spectral energy distribution of PB 8, dust grains with a dust-to-gas ratio of 0.01 (by mass) were also included. It is found that the presence of metal-rich inclusions can explain the heavy element ORLs, while a dual-dust chemistry with different grain species and discrete grain sizes likely produces the infrared continuum of this PN. This study demonstrates that the bi-abundance hypothesis, which was examined in a few PNe with large abundance discrepancies (ADFs\,$>$\,10), could also be applied to those typical PNe with moderate abundance discrepancies.
\end{abstract}

\end{frontmatter}
\section{Introduction}
\label{pb8:sec:introduction}

The planetary nebula (PN) PB\,8 (PN\,G292.4$+$04.1) has been the subject of some recent studies
\citep{Garcia-Rojas2009,Todt2010,MillerBertolami2011}. The central star of PB\,8 has been classified as [WC\,5-6] by \citet{Acker2003}, as weak emission-line star type \citep[\textit{wels};][]{Tylenda1993,Gesicki2006}, as [WC]-PG\,1159  \citep{Parthasarathy1998}, and finally as [WN/WC] hybrid by \citet{Todt2010}. Particularly, it is one of the rare stars, which has provoked a lot controversy about its evolutionary origin \citep{MillerBertolami2011}.
A detailed abundance analysis of the nebula by \citet{Garcia-Rojas2009} showed abundance discrepancy factors (ADF\,$\equiv$\,ORL/CEL) of $2.57$ for the O$^{++}$ ion and $1.94$ for the N$^{++}$ ion, which are in the range of typical ADFs observed in PNe \citep[ADFs\,$\sim$\,1.6--3.2; see review by][]{Liu2006a}. The nebular morphology was described as a round nebula with inner knots or filaments by \citet{Stanghellini1993}, and classified as elliptical by \citet{Gorny1997}. However, a narrow-band H$\alpha$+[N\,{\sc ii}] image of PB\,8 taken by \citet{Schwarz1992} shows a roughly spherical nebula with an angular diameter of about 7 arcsec \citep[6.5 arcsec $\times$ 6.6 arcsec reported by][]{Tylenda2003}, which is used throughout this paper.

\begin{table*}[t]
\footnotesize
\begin{center}
\caption{Journal of the Observations for PB~8.}
\begin{tabular}{llcccll}
\hline
\hline
Observatory & Obs Date  & $\lambda$-range({\AA}) & FWHM({\AA}) & Inst./Mod.& Program ID/PI  &Exp.Time (s) \\
\hline
Magellan 6.5-m & 2006 May 9  & 3350--5050, 4950--9400 & 0.15, 0.25 & MIKE &  M. Pe{\~n}a & 300,\,600,\,900  \\
\noalign{\vskip2pt}
ANU 2.3-m & 2010 Apr. 21  & 4415--5589, 5222--7070 & 0.83, 1.03 & WiFeS & 1100147, Q.A. Parker & 60,\,1200 \\
\noalign{\vskip2pt}
\textit{Spitzer} & 2008 Feb. 25 & 5.2--14.5\,$\mu$m, 14--38\,$\mu$m & -- & SL, LL & 40115, G. Fazio & -- \\
\hline \label{pb8:tab:obs:journal}
\end{tabular}
\end{center}
\end{table*}

The ionic abundances of heavy elements derived from optical recombination lines (ORLs) have been found to be systematically higher than those derived from collisionally excited lines (CELs) in many PNe \citep[see e.g.][]{Rola1994,Liu2000,Liu2001,Liu2006,Tsamis2004,Tsamis2008,Garcia-Rojas2009}. To solve this problem, \citet{Liu2000} suggested a bi-abundance model in which the nebula contains two components of different abundances: cold hydrogen-deficient `metal-rich' component and diffuse warm component of `normal' abundances. The H-deficient inclusions embedded in the nebular gas of normal abundances can dominate the emissions of ORLs \citep{Liu2000,Liu2004b}. The bi-abundance photoionization model of Abell~30 by \citet{Ercolano2003b} showed the possibility of such a scenario. More recently, the bi-abundance model by \citet{Yuan2011} was able to predict the ORLs in NGC 6153. Previously, the analysis of the emission-line spectrum of NGC 6153 by \citet{Liu2000} pointed to a component of the ionized gas, cold and very metal-rich. The photoionization modeling of NGC 1501 \citep{Ercolano2004} and Abell 48 \citep{Danehkar2014a} also suggested that they may contain some cold H-deficient structures.  

The aim of this paper is to construct photoionization models of PB~8, for which high quality spectroscopy has become available \citep{Garcia-Rojas2009}, constrained by an ionizing source determined using the Potsdam Wolf-Rayet (PoWR) models for expanding atmospheres \citep{Todt2010}. To reproduce the observed ORLs, a bi-abundance model is used, which consists of a chemically homogeneous abundance distribution containing a small fraction of dense metal-rich structures. In addition, the dust properties are constrained using the \textit{Spitzer} infrared (IR) continuum of the PN PB~8. The observations and modeling procedure are respectively described in Sections~\ref{pb8:sec:observations} and \ref{pb8:sec:photoionization}. In Section~\ref{pb8:sec:results}, we present our modeling results, while our conclusions are given in Section~\ref{pb8:sec:conclusions}.

\section{Observations}
\label{pb8:sec:observations}

\begin{figure*}
\begin{center}
\includegraphics[width=6.0in]{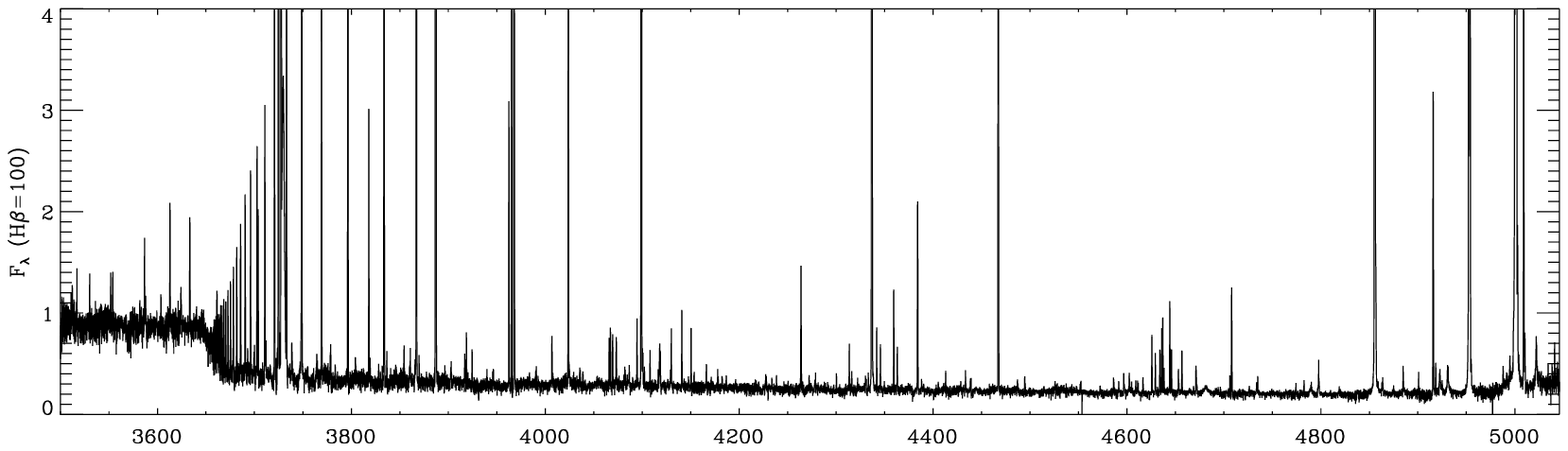}\\
\includegraphics[width=6.0in]{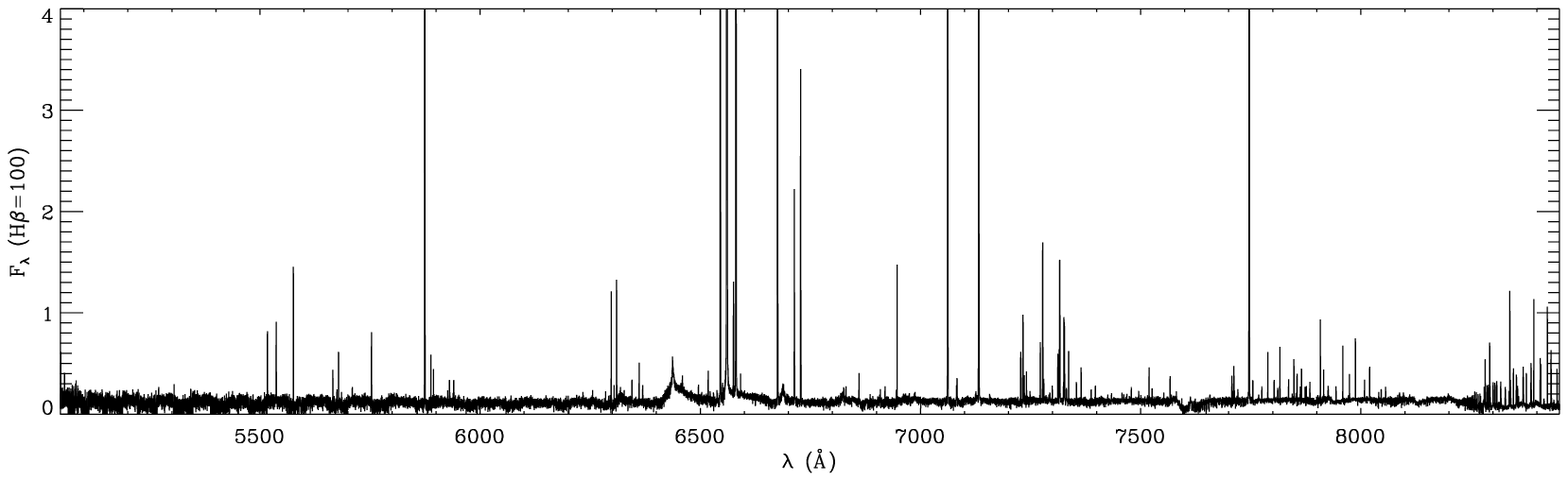}
\caption{The observed optical spectrum of the PN~PB~8 \citep{Garcia-Rojas2009}, covering wavelengths of (top) 3500--5046 {\AA} and (bottom) 5047--8451 {\AA}, and normalized such that $F$(H$\beta$)~=~100.}
\label{pb8:spectrum}
\end{center}
\end{figure*}

Deep optical long-slit spectra of the PN PB\,8 were obtained at Las Campanas Observatory (PI: M. Pe{\~n}a), using the 6.5-m Magellan telescope and the double echelle Magellan Inamori Kyocera Echelle (MIKE) spectrograph on 9 May 2006 \citep[][]{Garcia-Rojas2009}. An observational journal is presented in Table~\ref{pb8:tab:obs:journal}. The standard grating settings used yield wavelength coverage from 3350-5050\,{\AA} in the blue and 4950-9400\,{\AA} in the red. The mean spectral resolution is  0.15\,{\AA} FWHM in the blue and 0.25\,{\AA} FWHM in the red. The MIKE observations were taken with three individual exposures of 300, 600 and 900 sec using a slit of $1 \times 5$ arcsec$^2$ and a position angle (PA) of 345$^{\circ}$ passing through the central star. To prevent contamination of the stellar continuum, an area of $0.9 \times 1$ arcsec$^2$ on a bright knot located in the northern part of the slit was used to extract the nebular spectrum.
However, there is no definite constraint on the location of the combined slit spectrum for the bright knot in the nebula, as the slit crossed over the nebula during the three different observations.\footnote{The MIKE spectrograph was not attached to any telescope rotator in 2006, so the slit passes over the nebula in each observation (Garc{\'{\i}}a-Rojas, private communications, 2014).}   
The top and bottom panels of Figure~\ref{pb8:spectrum} show the blue and red spectra of PB\,8 extracted from the 2D MIKE echellograms, normalized such that $F$(H$\beta$)~=~100. As seen, several recombination lines from heavy element ions have been observed.  

\begin{table}
\footnotesize
\begin{center}
\caption{IR line fluxes of the PN PB\,8.}
\begin{tabular}{lcc}
\hline
\hline
Lines   & $F(\lambda)$ & $I(\lambda)$ \\
        & $10^{-12}$ erg\,cm$^{-2}$\,s$^{-1}$ & [$I($H$\beta)=100$] \\
\hline
 \fariii\ 8.99 $\mu$m   & 2.95  & 14.97 \\
 \fneii\ 12.82 $\mu$m   & 4.80  &  24.37 \\
 \fneiii\ 15.55 $\mu$m      & 21.60 & 110.66 \\
 \fsiii\ 18.68 $\mu$m   & 10.80 & 54.82 \\
 \fsiii\ 33.65 $\mu$m   & 5.98  & 30.36 \\
 \fneiii\ 36.02 $\mu$m      & 1.45  & 7.36 \\
\hline \label{pb8:tab:irs:flux}
\end{tabular}
\end{center}
\begin{list}{}{}
\item[\textbf{Note.}]Fig.~\ref{pb8:SED:Spitzer} shows the \textit{Spitzer} spectrum (SL \& LL combined).
\end{list}
\end{table}

Infrared (IR) spectra of the PN PB\,8 were taken on 25 February 2008 with the IR spectrograph on board the \textit{Spitzer} Space Telescope (programme ID 40115, PI: Giovanni Fazio). The flux calibrated IR spectra used in this paper have been obtained from the Cornell Atlas of \textit{Spitzer} / Infrared Spectrograph Sources\footnote{The Cornell Atlas of Spitzer/IRS Sources (CASSIS) is a product of the Infrared Science Center at Cornell University, supported by NASA and JPL. Website: http://cassis.astro.cornell.edu} \citep[CASSIS;][]{Lebouteiller2011,Lebouteiller2015}. The \textit{Spitzer} observations were taken with two low-resolution modules: Short--Low (SL) and Long--Low (LL). The SL spectrum was taken with an aperture size of $3.7 \times 57$ arcsec$^2$ covering a wavelength coverage of 5.2-14.5\,$\mu$m, whereas the LL spectrum has a wavelength coverage of 14.0-38.0\,$\mu$m and an aperture size of $10.7 \times 168$ arcsec$^2$. As the LL aperture is larger than the SL aperture, the LL module collects more flux than the SL, including the surrounding background  contamination. This causes a jump at around 14\,$\mu$m between the SL and the LL. To correct it, the LL spectrum was scaled to match the SL spectrum, so the combined \textit{Spitzer} spectrum describes the thermal IR emission of the nebula with little background contribution. Table~\ref{pb8:tab:irs:flux} lists the line fluxes measured from  the \textit{Spitzer} IR spectra (see Fig.~\ref{pb8:SED:Spitzer} for the \textit{Spitzer} SL and LL combined spectrum). The intrinsic line fluxes presented in column 3 are on a scale where $I($H$\beta)=100$, and the dereddened flux $I($H$\beta)=19.7 \times 10^{-12}$ erg\,cm$^{-2}$\,s$^{-1}$ calculated using the total H$\alpha$ flux \citep{Frew2013}, $E(B-V)=0.41$ and $R_{V} = 4$ \citep{Todt2010}.

Integral field unit (IFU) spectra were obtained at the Siding Spring Observatory on 21 April 2010 (programme ID 1100147, PI: Q.A. Parker), using the 2.3-m ANU telescope and the Wide Field Spectrograph \citep[WiFeS;][]{Dopita2007,Dopita2010}. The settings used were the B7000/R7000 grating combination and the RT\,560 dichroic, giving wavelength coverage from 4415-5589\,{\AA} in the blue and 5222-7070\,{\AA} in the red, and mean spectral resolution of 0.83\,{\AA} FWHM in the blue and 1.03\,{\AA} FWHM in the red (see the observational journal presented in Table~\ref{pb8:tab:obs:journal}). The WiFeS IFU rawdata were reduced using the \textsc{iraf} pipeline \textsf{wifes}, which consists of bias-reduction, sky-subtraction, flat-fielding, wavelength calibration using Cu-Ar arc exposures, spatial calibration using wire frames, differential atmospheric refraction correction, and flux calibration using spectrophotometric standard stars EG\,274 and LTT\,3864 \citep[fully described in][]{Danehkar2013a,Danehkar2014a}.

Figure~\ref{pb8:fig4} shows the spatially resolved flux intensity and radial velocity maps of PB\,8 extracted from the emission line $[$N\,{\sc ii}$]$ $\lambda$6584 for spaxels across the WiFeS IFU field.
The black/white contour lines depict the distribution of the emission of H$\alpha$ obtained from the SuperCOSMOS H$\alpha$ Sky Survey \citep[SHS;][]{Parker2005}, which can aid us in distinguishing the nebular borders. The emission line maps were obtained from solutions of the nonlinear least-squares minimization to a Gaussian curve function for each spaxel. The observed velocity $V_{\rm obs}$ was transferred to the local standard of rest (LSR) radial velocity $V_{\rm LSR}$. The WiFeS IFU observations have recently been used for morpho-kinematic studies of PNe \citep{Danehkar2015,Danehkar2016}. Considering the spatial resolution of the WiFeS (1 arcsec), this PN is very compact for detailed morpho-kinematic modeling. Following \citet{Danehkar2015a}, the tenuous lobes of PB~8 extending from its compact core can be used to determine its spatial orientation.  As seen in Figure~\ref{pb8:fig4}, the orientation of its faint lobes onto the plane of the sky has a position angle of $132^{\circ}\pm8^{\circ}$ relative to the north equatorial pole towards the east. Transferring into the Galactic coordinate system, its symmetric axis has a Galactic position angle of $114.6^{\circ}\pm8^{\circ}$, measured from the north Galactic pole towards the Galactic east, approximately aligned with the Galactic plane. 

We obtained an expansion velocity of $V_{\rm exp}=20\pm 4$\,km\,s$^{-1}$  from the $[$N\,{\sc ii}$]$ $\lambda$6584 flux integrated across the whole nebula in the WiFeS field, which is in agreement with $V_{\rm exp}=19$\,km\,s$^{-1}$ from $[$N\,{\sc ii}$]$ $\lambda$6584 line derived by \citet{Todt2010}. Moreover, \citet{Garcia-Rojas2009}  derived an expansion velocity of $V_{\rm exp}=14 \pm 2$\,km\,s$^{-1}$ from the $[$O\,{\sc iii}$]$ $\lambda$5007 line, which is associated with a different ionization zone. \citet{Garcia-Rojas2009} obtained $V_{\rm sys}=1.4$\,km\,s$^{-1}$ from $[$O\,{\sc iii}$]$ lines, in agreement with the value of $V_{\rm sys}=2.4$\,km\,s$^{-1}$ given by {\citet{Todt2010}.  

\begin{figure}
\begin{center}
\includegraphics[width=1.6in]{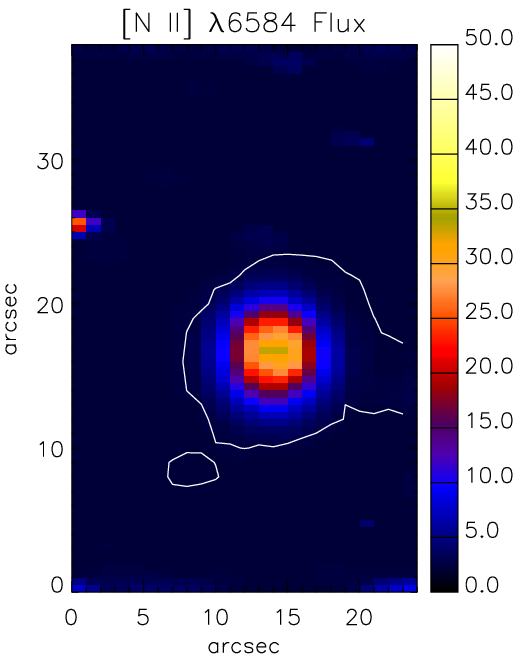}%
\includegraphics[width=1.6in]{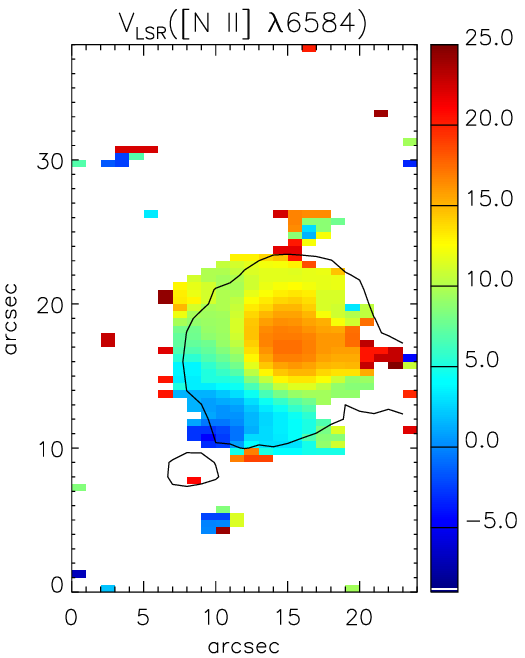}%
\caption{Maps of PB~8 in $[$N\,{\sc ii}$]$ $\lambda$6584 from the IFU observation. From
left to right: spatial distribution maps of flux intensity and LSR velocity. Flux unit is in $10^{-15}$~erg\,s${}^{-1}$\,cm${}^{-2}$\,spaxel${}^{-1}$, and velocity in km\,s${}^{-1}$. North is up and east is toward the left-hand side.
The white/black contour lines show the distribution of the narrow-band emission of H$\alpha$ in arbitrary unit obtained from the SHS \citep{Parker2005}.
\label{pb8:fig4}%
}%
\end{center}
\end{figure}

\section{Photoionization Modeling}
\label{pb8:sec:photoionization}

The photoionization modeling is performed using the {\sc mocassin} code (version 2.02.70), described in detail by \citet{Ercolano2003,Ercolano2005} in which the radiative transfer of the stellar and
diffuse field is computed using a Monte Carlo (MC) method constructed in a cubical Cartesian grid, allowing completely arbitrary distribution of nebular density and chemical abundances. This code has already been used to study some chemical inhomogeneous models, namely the H-deficient knots of Abell~30 \citep{Ercolano2003b} and the super-metal-rich knots of NGC~6153 \citep{Yuan2011}. To solve the problem of ORL-CEL abundance discrepancies in those PNe, they used a metal-rich component, whose ratios of heavy elements with respect to H are higher than those of the normal component.

To investigate the abundance discrepancies between the ORLs and CELs, we have constructed different photoionization models for PB~8. We run a set of model simulations, from which we finally selected  three models, which best reproduced the observations. Our first model (MC1) consists of a chemically homogeneous abundance distribution. Our second model (MC2) is roughly similar, but it includes some dense metal-rich knots (cells) embedded in the density model of normal abundances (see Section\,\ref{pb8:sec:density:model}). The final model (MC3) includes dust grains to match the \textit{Spitzer} IR observation (see Section\,\ref{pb8:sec:dust:model}). The atomic data sets used for our models include energy levels, collision strengths, and transition probabilities from Version 7.0 of the CHIANTI database \citep[][and references therein]{Landi2012}, hydrogen and helium free--bound coefficients of \citet{Ercolano2006}, and opacities from \citet{Verner1993} and \citet{Verner1995}.

The model parameters, as well as the physical properties for the models, are summarized in Table~\ref{pb8:modelparameters}, and discussed in more detail in the following sections. 
The modeling procedure consists of an iterative process, involving the comparison of the predicted emission line fluxes with the values measured from the observations, and the ionization and thermal structures with the values derived from the empirical analysis. 
The free parameters used in our models should be the nebular abundances, as the nebular density is adopted based on empirical results \citep{Garcia-Rojas2009}, and the stellar parameters based on the model atmosphere study \citep{Todt2010}. However, it is impossible to isolate effects of any parameter from each other, as they are dependent on each other, so we cannot just modify the nebular abundances without slightly adjusting the density distribution and the distance. The nebular ionization structure depends on the gas density and the stellar characteristics, so we fairly adjusted them to obtain the best-fitting models. We adopted the effective temperature of $T_{\rm eff}=52$\,kK, stellar luminosity of $L_{\star}=6000{\rm L}_{\odot}$, and non-local thermodynamic equilibrium (non-LTE) model atmosphere determined by \citet{Todt2010}. 
The optical depth for Lyman-continuum radiation of $\tau($Ly-c$)=0.63$ estimated by \citet{Lenzuni1989} indicated some ionizing radiation fields may escape from the nebular shell, so it could be a matter-bounded PN. Therefore, we attempted to adjust distance and gas density to reproduce the nebular total H$\beta$ intrinsic line flux. It is found that a model with an electron density of about $2550\pm550$~cm$^{-3}$ empirically derived by \citet{Garcia-Rojas2009} can well reproduce the nebular H$\beta$ intrinsic line flux at a distance of 4.9~kpc. 
We initially used the elemental abundances determined by \citet{Garcia-Rojas2009}, but we  adjusted them to match the observed nebular spectrum.

\begin{table*}
\footnotesize
\centering
\caption{Model parameters and physical properties for the final photoionization models.
\label{pb8:modelparameters}
}
\begin{tabular}{lcc|c|ccc|ccc}
\noalign{\vskip3pt} \noalign{\hrule} \noalign{\vskip3pt}
                                         &\multicolumn{2}{c|}{Empirical}&\multicolumn{7}{c}{Models} \\
                                         &\multicolumn{2}{c|}{ }& MC1       &\multicolumn{3}{c|}{MC2} &\multicolumn{3}{c}{MC3} \\
Parameter                                 &CEL         &ORL        &          &Normal        &Metal-rich&Total   &Normal&Metal-rich&Total      \\
\noalign{\vskip3pt} \noalign{\hrule}\noalign{\vskip3pt}
\teff\ (kK)                               &\multicolumn{2}{c|}{52}  &52        &\multicolumn{3}{c|}{52}  &\multicolumn{3}{c}{52}  \\
$L_{\star}$ (L$_{\odot}$)              &\multicolumn{2}{c|}{6000}  &6000      &\multicolumn{3}{c|}{6000}&\multicolumn{3}{c}{6000}\\
    \noalign{\vskip3pt}
$R_{\rm in}$ ($10^{17}$ cm)               &\multicolumn{2}{c|}{--}  & 0.8      &\multicolumn{3}{c|}{0.8} &\multicolumn{3}{c}{0.8} \\
$R_{\rm out}$ ($10^{17}$ cm)              &\multicolumn{2}{c|}{--}  & 2.6      &\multicolumn{3}{c|}{2.6} &\multicolumn{3}{c}{2.6} \\
    \noalign{\vskip3pt}
Filling factor                            &\multicolumn{2}{c|}{--}  & 1.000    & 0.944       & 0.056  & 1.000 & 0.944    & 0.056  & 1.000 \\
$\langle$\hyd$\rangle$ (\cmt)             &\multicolumn{2}{c|}{--}  & 2009    & 1957       & 3300     & 2032  & 1957    & 3300  &  2032 \\
$\langle$\eld$\rangle$ (\cmt)             &\multicolumn{2}{c|}{$2550\pm550$}       & 2257     & 2199       &4012      &  2301 & 2199    & 4012  &  2301 \\
$\rho_{\rm d}/\rho_{\rm g}$                         &\multicolumn{2}{c|}{--}  & --      &{ }&{ }&{--} &{ }&{ }&{0.01} \\
   \noalign{\vskip2pt}
He/H                                      &--         &0.122       &0.122     &0.122       &0.20      &   0.129   &0.122    &0.20   &   0.129   \\
C/H \, \ \ $\times$10$^5$                 &--         &72.25       & 63.0     &63.0        & 63.0     &  63.0    &63.0     & 63.0  &    63.0  \\
   \noalign{\vskip2pt}
N/H \, \ \ $\times$10$^5$                &16.22      &31.41\,$^{\mathrm{a}}$       & 11.0     & 6.1        &298.0\,$^{\mathrm{a}}$      &  32.7\,$^{\mathrm{a}}$   & 6.1     &298.0\,$^{\mathrm{a}}$   &   32.7\,$^{\mathrm{a}}$    \\
O/H \, \ \ $\times$10$^5$               &57.54      &146.61\,$^{\mathrm{a}}$        & 40.0     & 58.7       &551.0\,$^{\mathrm{a}}$      &  103.5\,$^{\mathrm{a}}$    & 58.7    &551.0\,$^{\mathrm{a}}$   &   103.5\,$^{\mathrm{a}}$  \\
   \noalign{\vskip2pt}
Ne/H   \ $\times$10$^5$                   &13.49      &19.9\,$^{\mathrm{a}}$         & 10.0     &15.0      & 15.0      \,$^{\mathrm{a}}$  &   15.0\,$^{\mathrm{a}}$    &15.0      & 15.0\,$^{\mathrm{a}}$   &   15.0\,$^{\mathrm{a}}$    \\
S/H \, \ \ $\times$10$^7$                 &204.17     &--          & 300.0    &300.0     &300.0       & 300.0  &300.0     &300.0  &  300.0 \\
Cl/H   \ $\times$10$^7$                   &2.0        &--          & 1.2      &1.6       &1.6         &  1.6  &1.6       &1.6   &  1.6  \\
Ar/H  \ $\times$10$^7$                    &43.65      &--          & 39.0     &45.0      &45.0        &  45.0   &45.0      &45.0  &  45.0   \\
   \noalign{\vskip3pt} \noalign{\hrule}\noalign{\vskip3pt}
\end{tabular}
\begin{list}{}{}
\item[$^{\mathrm{a}}$]The ORL empirical abundances were calculated from the ORLs over the total H$^{+}$ emission flux, emitted from both the diffuse gas and metal-rich inclusion, so the empirical abundances of the ORLs cannot be the same as the model metal-rich component, but roughly similar to the mean total abundances of both the metal-rich and normal components. 
\end{list}
\end{table*}

\subsection{The ionizing spectrum}
\label{pb8:sec:cspn:model}

The H-deficient non-LTE model atmosphere \citep{Todt2010}, which was calculated using the PoWR models for expanding atmospheres \citep{Grafener2002,Hamann2004}, was used as an ionizing source in our photoionization models. The PoWR models were constructed by solving the non-LTE radiative transfer equation of an expanding stellar atmosphere under the assumptions of spherical symmetry and chemical homogeneity. 
The PoWR model used was calculated for the stellar surface abundances H:He:C:N:O~=~40:55:1.3:2:1.3 (by mass), the stellar temperature $T_{\rm eff}$\,=\,52\,kK, the stellar luminosity $L_{\star}=6000{\rm L}_{\odot}$, the transformed radius $\log R_{\rm t}=1.43$\,R${}_{\odot}$ and the wind terminal velocity $V_{\infty}=1000$~km\,s$^{-1}$, which well matches the dereddened stellar spectra from FUSE, IUE and MIKE, as well as 2MASS JHK bands \citep{Todt2010}. We see that the nebular $[$O\,{\sc iii}$]$ $\lambda$5007 line flux relative to the H$\beta$ flux is well reproduced with an effective temperature of $T_{\rm eff}=52$\,kK in our photoionization models. The stellar luminosity of $L_{\star}=6000{\rm L}_{\odot}$ adopted by \citet{Todt2010} is related to a remnant core with a typical mass of $0.6M_{\odot}$ \citep[e.g.][]{MillerBertolami2007,Schonberner2005a}. 
The distance was also varied in order to reproduce the nebular emission-line fluxes, under the constraints of our adopted stellar parameters and spherical density distribution. The best results for the photoionization models were obtained at a distance of 4.9~kpc. 

Figure~\ref{pb8:SED:PoWR} compares the non-LTE model atmosphere flux of PB~8 with a blackbody flux at the same temperature. At energies higher than 54~eV (He\,{\sc ii} ground state), there is a significant difference between the non-LTE model atmosphere and blackbody flux. As discussed by \citet{Rauch2003}, a blackbody is not an accurate representation of the ionizing flux. The H-deficient non-LTE model atmosphere has a major departure from the solar model atmosphere at higher energies due to the small opacity from hydrogen. In our photoionization models, we theretofore used an non-LTE model atmosphere as the ionizing source to provide the best fit to the nebular spectrum. 
However, the difference may not be largely noticeable in our model as high-excitation lines (e.g. He\,{\sc ii}) are not observed.

\subsection{The density distribution}
\label{pb8:sec:density:model}

The initial nebular model to be run was the simplest possible density distribution, a homogeneous spherical geometry, to reproduce the CELs. The chemical abundances were taken to be homogeneous. A first attempt was made by using a homogeneous density distribution of 2550~cm$^{-3}$ with different inner and outer radii, deduced from the [O\,{\sc ii}], [S\,{\sc ii}] and [Cl\,{\sc iii}] lines \citep{Garcia-Rojas2009}. However, the uniform density distribution did not match the ionization and thermal structures, so we examined different density distributions adopted based on radiation-hydrodynamics simulations \citep[see e.g.][]{Perinotto2004,Schonberner2005a,Schonberner2005b} 
 to make the best fit to the observed CELs, and also to constrain the shell thickness. Some radiation-hydrodynamics results depict a radial density profile having the form $N_{\rm H}(r) = N_{0}[(r_{\rm out}/r)^{\alpha}]$, where $r$ is the radial distance from the center, $\alpha$ the radial density dependence, $N_{0}$ the characteristic density, $r_{\rm out}$ the outer radius. 
We finally adopted a density distribution with a powerlaw radial profile. Figure \ref{pb8:density:model2} illustrates the 3-D spherical density model constructed based on the radial density profile, and the ionizing source is located in the corner. 
We chose the characteristic density of $N_{0}=2600$\,cm$^{-3}$ and the radial density dependence of $\alpha=-1$.  The outer radius of the sphere is equal to $R_{\rm out} = 3.5$ arcsec and the thickness is $\delta R =2.4$ arcsec. 
We examined distances, with values within the range of distances 2.2 and 5.8 kpc \citep[][and references therein]{Phillips2004}. The distance of $D=4.9$ kpc found here, was chosen, because of the best predicted H$\beta$ luminosity $L({\rm H}\beta)=4\pi D^2 I({\rm H}\beta)$, and it is within the range of distances 4.2 and 5.15 kpc estimated by \citet{Todt2010}. Taking the angular diameter of 7 arcsec, we derive an outer radios of $R_{\rm out}=2.6 \times 10^{17}$ cm at the given distance of $D=4.9$ kpc.

\begin{figure}
\begin{center}
\includegraphics[width=3.3in]{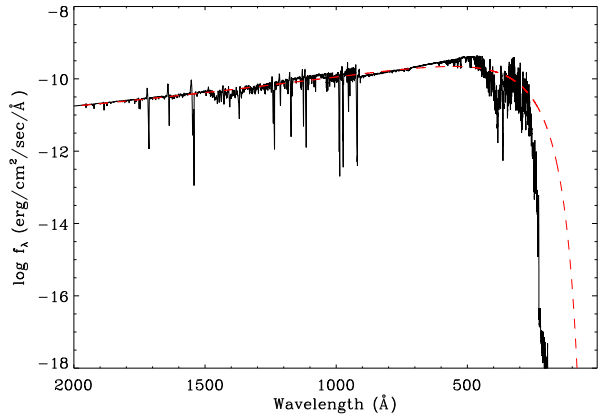}%
\caption{Non-LTE model atmosphere (solid line) calculated with $T_{\rm eff}$\,=\,52\,kK and chemical abundance ratio of  H:He:C:N:O~=~40:55:1.3:2:1.3 by mass \citep{Todt2010}, compared with a blackbody (dashed line) at the same temperature.
}%
\label{pb8:SED:PoWR}%
\end{center}
\end{figure}

\begin{table}
\footnotesize
\begin{center}
\caption{Metal-rich component parameters in the model MC2.}
\begin{tabular}{lc}
\hline
\hline
Parameter & Value  \\
\hline
Filling factor &  0.056 \\
Mass fraction  &  6.3 percent \\
Number of knots  & 33   \\
Knot size (arcsec$^{3}$)  &  $(0.35)^{3}$    \\
$\langle$\hyd$\rangle$ (\cmt)  & 3300 \\
$\langle$\eld$\rangle$ (\cmt)  & 4012 \\
$\langle$\elt$\rangle$ (K)  & 4286 \\
\hline \label{pb8:tab:hpoor:model}
\end{tabular}
\end{center}
\end{table}

A second attempt (MC2) was to reproduce the observed ORLs by introducing a fraction of metal-rich knots into the density distribution used by the first model. The abundance ratios of He, N, and O relative to H in the metal-rich component are higher than those in the normal component. Two different bi-abundance models suggested by \citet{Liu2000}: one assumes that H-deficient component with a very high density ($N_{\rm e} = 2 \times 10^{6}$\,cm$^{-3}$) and a moderate temperature ($T_{\rm e}= 4700$\,K), which has been adopted in the bi-abundance photoionization model of Abell~30 \citep{Ercolano2003b} and NGC 6153 \citep{Yuan2011}. However, H-deficient component in the second model introduced by \citet{Liu2000} has an 8 times lower $N_{\rm e}$ and a 20 times
lower $T_{\rm e}$ than the normal component. After exploring both the assumptions, we adopted a dense metal-rich component with the H number density of 3300\,cm$^{-3}$, which is roughly 1.7 times higher than the mean density of the surrounding material. 

In Figure\,\ref{pb8:density:model2}, the 3-D spatial distribution of metal-rich knots (cells) are shown as a darker shade inside the density model of normal abundances. The gas-filling factor in the density model MC1 was kept at unity, while the inclusion of 33 metal-rich cells in the normal abundances nebula in the model MC2 (see Table~\ref{pb8:tab:hpoor:model}) leads to gas-filling factors of 0.056 for the metal-rich component. It means that the dense metal-rich inclusions occupy 5.6 percent of the total gaseous volume of the nebula. Higher values of the gas-filling factor for the metal-rich component require lower values of the number density in order to reproduce the ORLs, but they may not provide the best match. We note that the filling factor parameter in {\sc mocassin} is still set to 1.0, which can have a major role in emission-line calculations. For example, reducing the {\sc mocassin} filling factor parameter from 1.0 to 0.5 decreases the nebular H$\beta$ luminosity by a factor of 2 and increases the $[$O\,{\sc iii}$]$ $\lambda$5007/H$\beta$ flux ratio by 9 percent, so the nebula cannot be reproduced. 

\begin{figure}
\begin{center}
\includegraphics[width=3.0in]{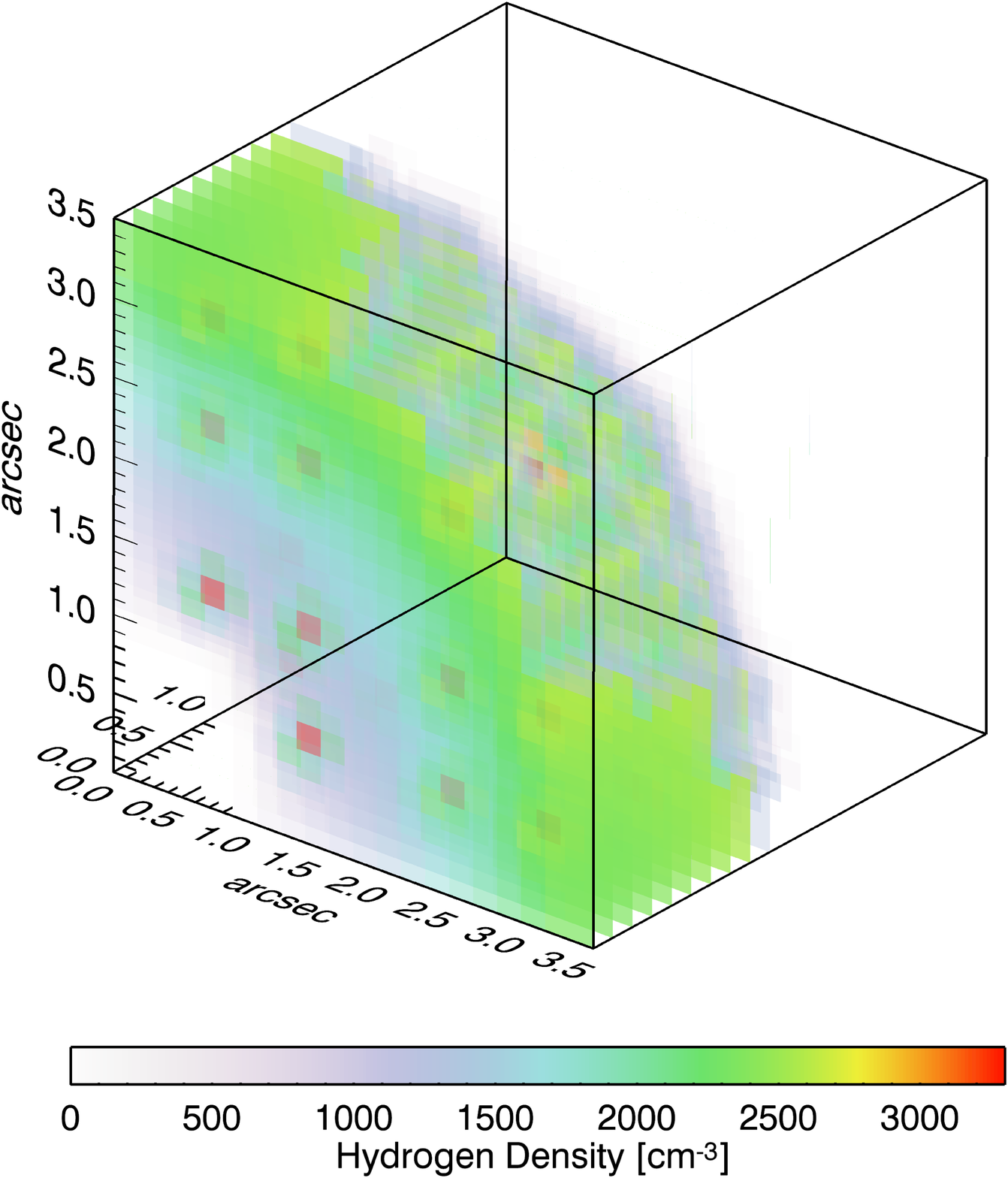}%
\caption{Spherical density distribution a powerlaw radial profile adopted for photoionization models. The sphere has outer radius of $3.5$ arcsec and thickness of $2.4$ arcsec. The ionizing source is placed in the corner (0, 0, 0). Metal-rich inclusions are shown as a darker knots with $N_{H}=3300$\,cm$^{-3}$ embedded in the density model of normal abundances. The units of the axes are arcsec.
}%
\label{pb8:density:model2}%
\end{center}
\end{figure}

For our bi-abundance model, we assume 33 metal-rich knots are uniformly distributed inside the diffuse warm nebula. However, \citet{Ercolano2003b} adopted a dense core surrounded by a less dense outer region in the bi-abundance model of Abell~30. Moreover, the metal-rich inclusions were distributed in the inner region of the diffuse warm nebula in the bi-abundance model of NGC~6153 \citep{Yuan2011}. Adopting the current knot density (3300\,cm$^{-3}$) in PB\,8, different distributions of metal-rich knots require totally different N/H and O/H abundance ratios and filling factors, which may not be in pressure equilibrium with their diffuse warm nebula. We note that ADFs of Abell~30 and NGC~6153 are very large, and are not similar to moderate ADFs of PB\,8. Moreover, the effective temperatures of the central stars of Abell~30 and NGC~6153 are 130\,000 and 90\,000\,K, respectively, which are hotter than $T_{\rm eff}=$\,52\,000\,K in PB\,8. This indicates that the central star of PB\,8 is in an early stage of its evolutionary path towards becoming a white dwarf, and should be younger than the hot central stars of Abell~30 and NGC~6153. 

The physical size of the metal-rich knots also play an important role in making an ionization structure to reproduce the \nii\ and \oii\ ORLs. It is found that the physical size of (0.008)$^3$\,parsec$^3$ (at $D=4.9$ kpc) can well produce the observed ORLs. Reducing the physical size will lead to an increase in the number of knots, which also need different density or/and N/H and O/H abundance ratios to match the observation. Smaller sizes will contribute to higher ionization and more thermal effects \citep[see also][]{Yuan2011}, so we may not be able to reproduce electron temperatures estimated from the ORLs (see Section\,\ref{pb8:sec:results:temp}). A larger physical size will also decreases the number of knots, but it needs different density or/and different elemental abundances of N and O, which may not make metal-rich inclusions in pressure equilibrium with the surrounding gas.

\subsection{The nebular elemental abundances}
\label{pb8:sec:abundances:model}

We used a homogeneous chemical abundance distribution for the model MC1 consisting of 9 elements, including all the major contributors to the thermal balance of the nebula and those producing the density- and temperature-sensitive CELs. The abundances derived from the empirical analysis \citep{Garcia-Rojas2009} were chosen as starting values; these were iteratively modified to get a better fit to the CELs. The final abundance values are listed in Table~\ref{pb8:modelparameters}, where they are given by number with respect to H.

A two-component  elemental abundance distribution was used for the model MC2 that yields a better fit to the observed ORLs. The parameters of the metal-rich cells included in the bi-abundance model MC2 are summarized in Table \ref{pb8:tab:hpoor:model}. The metal-rich inclusions were constructed using 33 knots with the physical size of $(0.35)^{3}$ arcsec$^{3}$. As shown in Figure\,\ref{pb8:density:model2} (b), they are uniformly distributed inside the normal component model with the same geometry, but a filling factor of 0.056. The initial guesses at the elemental abundances of N and O in the the metal-rich component were taken from the ORL empirical results; they were successively increased to fit the observed \nii\ and \oii\ ORLs. Table~\ref{pb8:modelparameters} lists the final elemental abundances (with respect to H) derived for both components, normal and metal-rich. The final model, which provided a better fit to most of the observed ORLs, has a total metal-rich mass of about 6.3 percent of the ionized mass of the entire nebula. 

The O/H and N/H abundance ratios in the metal-rich component are about 1.0 and 1.7 dex larger than those in the normal component.
The C/O abundance ratio in the metal-rich component less than unity is in disagreement with the theoretical predictions of born-again stellar models \citep{Herwig2001,Althaus2005,Werner2006}. 
As seen in Table~\ref{pb8:modelparameters}, the ORL total abundances empirically derived were not similar to the elemental abundances chosen for the model metal-rich component. This is due to the fact that the ORL empirical abundances were derived from the ORLs, emitted mainly from metal-rich inclusion, over the H$^{+}$ flux of the entire nebula, emitted from both the diffuse gas and metal-rich inclusion. Hence, the empirical abundances of the ORLs are roughly similar to the mean total abundances of both the metal-rich and normal components.

As seen in Table~\ref{pb8:modelparameters}, the elemental abundance for neon does not show a large abundance discrepancy similar to what we see for oxygen and nitrogen, which is unlike to the bi-abundance models of Abell~30 \citep{Ercolano2003b} and NGC~6153 \citep{Yuan2011}. We note that the H-deficient knots of Abell~30 shows ADF(Ne$^{++}$) values in the range of 400--1000 \citep{Wesson2003}, and NGC 6153 has a ADF(Ne$^{++}$) of about 60 \citep{Liu2000}. Nevertheless, PB\,8 has ADF(Ne$^{++}$)\,=\,1.48 \citep{Garcia-Rojas2009}. To reproduce the spectrum of Abell~30, \citet{Ercolano2003b} also assumed a bi-abundance model, in which the metal-rich core has a density of about six times higher than the surrounding normal envelope. Thus, both Abell~30 and NGC~6153 have extremely large ADFs, which are dissimilar to PB\,8. We should also include that the atomic data of the ORLs of Ne$^{++}$ ion \citep{Kisielius1998} used by {\sc mocassin} do not have any recombination coefficients for the \neii\ $\lambda$4391.94 and $\lambda$4409.30 lines, while \citet{Garcia-Rojas2009} employed different atomic data (Kisielius \& Storey; unpublished) to derive Ne$^{++}$ ion abundance from for \neii\ ORLs. 

It is worthwhile to mention that the AGB nucleosynthesis dramatically changes the composition of He, C and N \citep{Karakas2003,Karakas2009}, but other elements such as Ne, S, Cl, and Ar are left untouched by the evolution and nucleosynthesis in low and intermediate-mass stars. In this typical PN with moderate ADFs, abundances of other elements heavier than oxygen such as neon in the metal-rich components seem to be the same as those in the normal component.

\begin{table}
\footnotesize
\begin{center}
\caption{Input parameters for the dust model of PB\,8.}
\begin{tabular}{lll}
\hline
\hline
Grain Species & Weight &  Ref. for optical constants\\
\hline
Amorphous Carbon &   1 &  \citet{Hanner1988} \\
Crystalline Silicate   &  1 & \citet{Jaeger1994} \\
\hline 
Grain Radius ($\mu$m) & \multicolumn{2}{l}{Weight}  \\
\hline
$0.16$  & \multicolumn{2}{l}{50}    \\
$0.40$  & \multicolumn{2}{l}{1}  \\
\hline
\label{pb8:tab:dust:model}
\end{tabular}
\end{center}
\end{table}

\subsection{Dust modeling}
\label{pb8:sec:dust:model}

PB 8 is known to be very dusty \citep[e.g.][]{Lenzuni1989,Stasinska1999}, which must influence the radiative processes in the nebula. \citet{Lenzuni1989} studied the IRAS measurements (25, 60 and 100 $\mu$m fluxes), and derived a dust temperature of $T_{\rm d} = 85 \pm  0.4$\,K, an optical depth of $\tau($Ly-c$) = 0.63$ and a dust-to-gas mass ratio of $\rho_{\rm d}/\rho_{\rm g} =0.0123$ from a blackbody function fitted to the IRAS data. Similarly, \citet{Stasinska1999} determined $T_{\rm d} = 85$\,K, but $\rho_{\rm d}/\rho_{\rm g} =0.0096$ from the broad band IRAS data. 
From the comparison of the mid-IR emission with a blackbody model of $150$\,K, \citet{Todt2010} suggested that it possibly contains a warm dust with different dust compositions. We notice that the models MC1 and MC2 cannot provide thermal effects to account for the \textit{Spitzer} IR continuum, so a dust component is necessary to reproduce the spectral energy distribution (SED) of the nebula observed in the IR range. The third model (MC3) presented here treats dust properties of PB 8 using the dust radiative transfer features included in the {\sc mocassin} \citep{Ercolano2005}. Discrete grain sizes have been chosen based on the size range given by \citet{Mathis1977}. The absence of the 9.7 $\mu$m  amorphous silicate feature in the IR spectrum of PB 8 is commonly observed in O-rich circumstellar envelopes, which could imply this PN has a carbon-based dust. However, the strong features at 23.5, 27.5, and 33.8 $\mu$m are mostly attributed to crystalline silicates \citep{Molster2002}. The features seen at 6.2, 7.7, 8.6, and 11.3 $\mu$m  are related to polycyclic aromatic hydrocarbons (PAHs) \citep{Garcia-Lario1999}, together with broad features at 21 and 30 $\mu$m corresponding to a mixed chemistry having both O-rich and C-rich dust grains. The far-IR emission fluxes at 65 and 90 $\mu$m \citep{Yamamura2010} could be related to relatively warm forsterite grains, which emit at a longer wavelength. The 65 $\mu$m emission may be related to a crystalline water-ice structure, although its presence cannot be confirmed at the moment. 

\begin{figure}
\begin{center}
\includegraphics[width=3.3in]{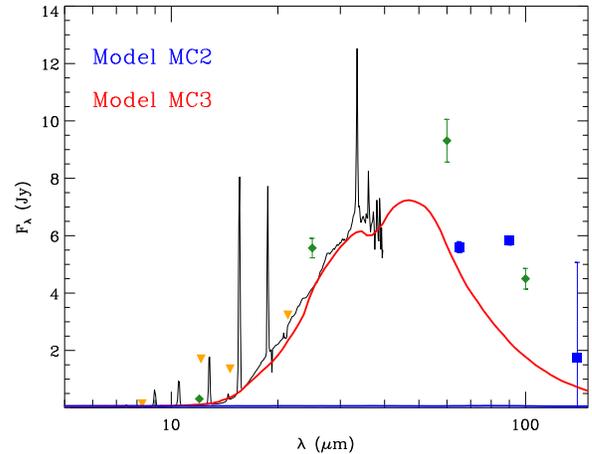}%
\caption{Observed \textit{Spitzer} spectrum (black line) of PB\,8 are compared with the continuum predicted by the model MC2 (blue line) and MC3 (red line). It also shows the photometric measurements for 12, 25, 60, and 100 $\mu$m (denoted by green diamonds) from IRAS \citep{Helou1988}, 8.3, 12.1, 14.7, and 21.34 $\mu$m (orange downward triangle) from MSX \citep{Egan2003}, and the far-IR measurements (blue squares) $F$(65\,$\mu$m$)=5.60\pm 0.19$, $F$(90\,$\mu$m$)=5.83\pm 0.16$, and $F$(140\,$\mu$m$)=1.74\pm 3.33$ Jy from AKARI/FIS \citep{Yamamura2010}. Note that the predicted nebular SED does not contain any nebular emission line fluxes.
}%
\label{pb8:SED:Spitzer}%
\end{center}
\end{figure}

The thermal IR emission of PB 8 was modeled by adding a mixed dust chemistry to the pure-gas photoionization model described in the previous sections. We explored a number of grain sizes and species, which could provide a best-fitting curve to the \textit{Spitzer} IR continuum (see Figure\,\ref{pb8:SED:Spitzer}). As seen, the model MC2 cannot produce the IR continuum, whereas the model MC3 fairly produces it. We tried to match the far-IR emission flux at 65 $\mu$m, while the 140 $\mu$m flux is extremely uncertain, $F$(140\,$\mu$m$)=1.74\pm 3.33$ Jy. The dust-to-gas mass ratio was varied until the best IR continuum flux was produced. Table~\ref{pb8:tab:dust:model} lists the dust parameters used for the final model of PB\,8, the dust-to-gas ratio is given in Table~\ref{pb8:modelparameters}. The geometry of the dust distribution is the same as the gas density distribution. The value of $\rho_{\rm d}/\rho_{\rm g} = 0.01$ found here is in agreement with \citet{Lenzuni1989}. The final dust model incorporates two different grains, amorphous carbon and crystalline silicate with optical constants taken from \citet{Hanner1988} and \citet{Jaeger1994}, respectively. We also note that the nebular SED (shown in Figure\,\ref{pb8:SED:Spitzer}), which is computed by {\sc mocassin},  does not contain 
any contributions from the nebular emission line fluxes. 

For PB\,8, \citet{Lenzuni1989} estimated a grain radius of $0.017$ $\mu$m from the thermal balance equation under the assumption of the UV absorption efficiency $Q_{\rm UV}=1$. \citet{Stasinska1999} argued that the method of \citeauthor{Lenzuni1989} underestimates the grain radius, and one cannot derive the grain size in such a way. Our photoionization modeling implies that dust grains with a radius of $0.017$ $\mu$m produce a very warm emission higher than $T_{\rm d} = 85$\,K. The final dust model uses two discrete grain sizes, namely grain radii of 0.16 $\mu$m (warm) and 0.40 $\mu$m (cool), which can fairly reproduce the observed thermal infrared SED with wavelengths less than 80 $\mu$m. Smaller grain sizes can produce hot emission that increases the continuum at shorter wavelengths ($<10$ $\mu$m), whereas larger grain sizes add cooler emission that may depict as a rise in the continuum at longer wavelengths ($>80$ $\mu$m). Although the current two grain sizes can well reproduce the \textit{Spitzer} SED of PB 8, the solutions may not be unique. A dust model with more than two grain sizes may also be possible, but it needs more computational simulations to find the best-fit model. Moreover, inhomogeneous dust distribution and viewing angles \citep[see Figure 2 in][]{Ercolano2005} can also change the predicted SED. As there is no information on the inclination of
dust grains and their geometry, so we assumed that they follow the gas density distribution. 

There is a discrepancy in fluxes with wavelengths higher than 80 $\mu$m, which could be attributed to a possible inhomogeneous dust distribution. We note that the band measurements with wavelengths higher 100 $\mu$m could have high uncertainties, e.g., $F$(140\,$\mu$m$)=1.74\pm 3.33$\,Jy. We also note a small discrepancy for fluxes with wavelengths less than 15 $\mu$m, which could be related to the difference between the SL aperture ($3.7 \times 57$ arcsec$^2$) and the LL aperture ($10.7 \times 168$ arcsec$^2$) used for the SL spectrum (5.2-14.5\,$\mu$m) and the LL spectrum (14.0-38.0\,$\mu$m), and uncertainties in scaling the LL spectrum (see Sec. \ref{pb8:sec:observations}). As both the LL and SL apertures covering some areas larger than the optical angular diameter of PB\,8 (7 arcsec), they could also be contaminated by the ISM surrounding the nebula.

\begin{table*}
\footnotesize
\centering
\caption{Comparison of predictions from the models and the observations.
The observed, dereddened intensities are in units such that $I$(\hb)$=100$.
Columns (6)--(12) give the ratios of predicted over observed values in each
case.
\label{pb8:modelresults}
}
\begin{tabular}{lrl|rr|r|rrr|rrr}
\noalign{\vskip3pt} \noalign{\hrule} \noalign{\vskip3pt}
Line    &$\lambda_{0}$(\AA) &Mult &$I_{\rm obs}$   &Err(\%)  &MC1     &\multicolumn{3}{c}{MC2}   &\multicolumn{3}{c}{MC3} \\
        &                   &     &                &                &        & Normal  & M-rich & Total & Normal  & M-rich & Total \\
\noalign{\vskip3pt} \noalign{\hrule}\noalign{\vskip3pt}
\multicolumn{10}{c}{H, He recombination lines}\\
     \hb& 4861.33&   H4&   100.000&     5.0&     1.000&     0.772&     0.228&     1.000&     0.771&     0.229&     1.000\\
     \ha& 6562.82&   H3&   282.564&     6.0&     1.031&     0.801&     0.247&     1.047&     0.799&     0.247&     1.047\\
     \hg& 4340.47&   H5&    45.666&     5.0&     1.019&     0.784&     0.228&     1.013&     0.784&     0.229&     1.013\\
     \hd& 4101.74&   H6&    24.285&     5.0&     1.057&     0.813&     0.235&     1.048&     0.813&     0.236&     1.049\\
     \hi& 3970.07&   H7&    14.466&     6.0&     1.089&     0.838&     0.242&     1.080&     0.837&     0.243&     1.080\\
     \hi& 3835.39&   H9&     6.784&     6.0&     1.069&     0.822&     0.238&     1.060&     0.822&     0.239&     1.061\\
\noalign{\vskip2pt}
    \hei& 3888.65&    2&    19.892&     6.0&     0.702&     0.531&     0.235&     0.766&     0.532&     0.236&     0.768\\
    \hei& 7065.28&   10&     4.265&     7.0&     0.752&     0.552&     0.202&     0.754&     0.555&     0.203&     0.758\\
    \hei& 5875.64&   11&    17.127&     6.0&     1.089&     0.852&     0.473&     1.325&     0.850&     0.473&     1.323\\
    \hei& 4471.47&   14&     6.476&     5.0&     1.014&     0.787&     0.411&     1.199&     0.786&     0.412&     1.198\\
    \hei& 4026.21&   18&     3.116&     6.0&     0.976&     0.757&     0.366&     1.123&     0.756&     0.367&     1.123\\
    \hei& 7281.35&   45&     0.815&     8.0&     1.126&     0.842&     0.339&     1.181&     0.844&     0.340&     1.185\\
    \hei& 6678.15&   46&     5.233&     6.0&     1.020&     0.802&     0.432&     1.233&     0.799&     0.432&     1.231\\
    \hei& 4921.93&   48&     1.737&     5.0&     1.014&     0.790&     0.402&     1.191&     0.788&     0.403&     1.191\\
\noalign{\vskip2pt}
\multicolumn{10}{c}{Heavy-element recombination lines}\\
    \cii& 6578.05&    2&     0.545&     9.0&     0.575&     0.438&     0.126&     0.563&     0.437&     0.126&     0.563\\
    \cii& 7231.34&    3&     0.234&    17.0&     1.088&     0.840&     0.257&     1.096&     0.837&     0.257&     1.094\\
    \cii& 7236.42&    3&     0.464&    10.0&     0.988&     0.762&     0.233&     0.995&     0.760&     0.233&     0.993\\
    \cii& 4267.15&    6&     0.781&     7.0&     0.857&     0.670&     0.218&     0.888&     0.667&     0.218&     0.885\\
\noalign{\vskip2pt}
    \nii& 5666.64&    3&     0.192&    25.0&     0.114&     0.048&     0.683&     0.731&     0.048&     0.685&     0.732\\
    \nii& 5676.02&    3&     0.084&       :&     0.116&     0.049&     0.693&     0.741&     0.048&     0.694&     0.743\\
    \nii& 5679.56&    3&     0.260&    18.0&     0.157&     0.066&     0.940&     1.006&     0.066&     0.942&     1.007\\
    \nii& 4601.48&    5&     0.099&    21.0&     0.073&     0.031&     0.420&     0.451&     0.030&     0.421&     0.452\\
    \nii& 4607.16&    5&     0.083&    25.0&     0.070&     0.029&     0.400&     0.429&     0.029&     0.401&     0.430\\
    \nii& 4613.87&    5&     0.063&    30.0&     0.069&     0.029&     0.395&     0.424&     0.029&     0.396&     0.424\\
    \nii& 4621.39&    5&     0.085&    24.0&     0.068&     0.028&     0.390&     0.418&     0.028&     0.391&     0.419\\
    \nii& 4630.54&    5&     0.289&    10.0&     0.075&     0.031&     0.429&     0.460&     0.031&     0.430&     0.461\\
    \nii& 4643.06&    5&     0.122&    18.0&     0.059&     0.025&     0.338&     0.362&     0.025&     0.339&     0.363\\
    \nii& 4994.37&   24&     0.099&    21.0&     0.066&     0.028&     0.420&     0.448&     0.028&     0.421&     0.449\\
    \nii& 5931.78&   28&     0.151&    30.0&     0.045&     0.019&     0.284&     0.303&     0.019&     0.285&     0.304\\
    \nii& 5941.65&   28&     0.115&       :&     0.110&     0.047&     0.696&     0.743&     0.046&     0.697&     0.743\\
\noalign{\vskip2pt}
    \oii& 4638.86&    1&     0.206&    12.0&     0.181&     0.197&     0.527&     0.724&     0.196&     0.527&     0.723\\
    \oii& 4641.81&    1&     0.380&     8.0&     0.248&     0.270&     0.720&     0.990&     0.268&     0.721&     0.989\\
    \oii& 4649.13&    1&     0.458&     8.0&     0.391&     0.426&     1.136&     1.562&     0.423&     1.137&     1.561\\
    \oii& 4650.84&    1&     0.221&    12.0&     0.169&     0.184&     0.491&     0.675&     0.183&     0.491&     0.674\\
    \oii& 4661.63&    1&     0.222&    12.0&     0.215&     0.234&     0.624&     0.858&     0.232&     0.625&     0.857\\
    \oii& 4676.24&    1&     0.184&    13.0&     0.218&     0.237&     0.633&     0.870&     0.236&     0.633&     0.869\\
    \oii& 4319.63&    2&     0.081&    26.0&     0.364&     0.397&     1.067&     1.465&     0.395&     1.068&     1.463\\
    \oii& 4336.83&    2&     0.054&    36.0&     0.161&     0.176&     0.472&     0.648&     0.175&     0.472&     0.647\\
    \oii& 4349.43&    2&     0.197&    13.0&     0.346&     0.378&     1.016&     1.395&     0.376&     1.017&     1.393\\
    \oii& 3749.48&    3&     0.281&    11.0&     0.132&     0.143&     0.375&     0.518&     0.142&     0.376&     0.518\\
    \oii& 4414.90&    5&     0.036&       :&     0.489&     0.524&     1.275&     1.799&     0.521&     1.278&     1.799\\
    \oii& 4416.97&    5&     0.090&    24.0&     0.109&     0.116&     0.283&     0.399&     0.116&     0.284&     0.399\\
    \oii& 4072.15&   10&     0.265&    11.0&     0.331&     0.363&     0.987&     1.350&     0.360&     0.988&     1.348\\
    \oii& 4075.86&   10&     0.275&    11.0&     0.460&     0.505&     1.374&     1.879&     0.501&     1.375&     1.876\\
    \oii& 4085.11&   10&     0.086&    26.0&     0.190&     0.209&     0.568&     0.776&     0.207&     0.568&     0.775\\
    \oii& 4121.46&   19&     0.163&    16.0&     0.063&     0.069&     0.191&     0.260&     0.069&     0.191&     0.260\\
    \oii& 4132.80&   19&     0.202&    13.0&     0.099&     0.109&     0.301&     0.410&     0.108&     0.301&     0.410\\
    \oii& 4153.30&   19&     0.250&    12.0&     0.115&     0.126&     0.348&     0.474&     0.125&     0.348&     0.473\\
    \oii& 4110.79&   20&     0.147&    17.0&     0.060&     0.066&     0.182&     0.248&     0.066&     0.182&     0.248\\
    \oii& 4119.22&   20&     0.087&    25.0&     0.374&     0.411&     1.133&     1.544&     0.408&     1.133&     1.541\\
    \oii& 4699.22&   25&     0.026&       :&     0.093&     0.103&     0.283&     0.386&     0.102&     0.283&     0.385\\
    \oii& 4906.81&   28&     0.096&    21.0&     0.096&     0.105&     0.289&     0.394&     0.104&     0.289&     0.394\\
    \oii& 4924.53&   28&     0.154&    15.0&     0.101&     0.111&     0.307&     0.418&     0.111&     0.307&     0.417\\
\noalign{\vskip3pt} \noalign{\hrule}\noalign{\vskip3pt}
\noalign{\vskip2pt}
\end{tabular}
\end{table*}

\begin{table*}[t]
\footnotesize
\centering
\begin{tabular}{lrl|rr|r|rrr|rrr}
\noalign{\vskip3pt} \noalign{\hrule} \noalign{\vskip3pt}
Line    &$\lambda_{0}$(\AA) &Mult &$I_{\rm obs}$   &Err(\%)  &MC1     &\multicolumn{3}{c}{MC2}   &\multicolumn{3}{c}{MC3} \\
        &                   &     &                &                &        & Normal  & M-rich & Total & Normal  & M-rich & Total \\
\noalign{\vskip3pt} \noalign{\hrule} \noalign{\vskip3pt}
\noalign{\vskip2pt}
\multicolumn{10}{c}{Collisionally excited lines}\\
  \fnii\,$^{\mathrm{d}}$ & 5754.64&   3F&     0.346&    14.0&     0.495&     0.186&     0.506&     0.692&     0.199&     0.509&     0.708\\
   \fnii& 6548.03&   1F&     7.667&     6.0&     0.926&     0.393&     0.613&     1.006&     0.418&     0.631&     1.048\\
   \fnii& 6583.41&   1F&    22.318&     6.0&     0.972&     0.412&     0.643&     1.055&     0.438&     0.662&     1.100\\
   \foii& 3726.03&   1F&    17.103&     6.0&     0.897&     0.947&     0.061&     1.008&     1.035&     0.065&     1.100\\
   \foii& 3728.82&   1F&     9.450&     6.0&     0.782&     0.813&     0.044&     0.857&     0.890&     0.047&     0.936\\
   \foii\,$^{\mathrm{d}}$& 7318.92&   2F&     0.227&    18.0&     0.530&     0.491&     0.538&     1.029&     0.546&     0.539&     1.085\\
   \foii\,$^{\mathrm{d}}$& 7319.99&   2F&     0.811&     8.0&     0.451&     0.418&     0.536&     0.954&     0.465&     0.538&     1.003\\
   \foii\,$^{\mathrm{d}}$& 7329.66&   2F&     0.387&    12.0&     0.518&     0.480&     0.537&     1.017&     0.535&     0.539&     1.074\\
   \foii\,$^{\mathrm{d}}$& 7330.73&   2F&     0.471&    10.0&     0.419&     0.388&     0.536&     0.924&     0.432&     0.537&     0.969\\
  \foiii& 4363.21&   2F&     0.528&     7.0&     1.733&     0.927&     0.007&     0.935&     0.995&     0.008&     1.003\\
  \foiii& 4958.91&   1F&   116.957&     5.0&     1.130&     0.858&     0.137&     0.995&     0.886&     0.143&     1.028\\
  \foiii& 5006.84&   1F&   348.532&     5.0&     1.132&     0.859&     0.137&     0.996&     0.887&     0.143&     1.030\\
\noalign{\vskip2pt}
  \fneii&   12.82$\mu$m&    ~&    24.370&  \ldots&     0.516&     0.696&     0.209&     0.905&     0.716&     0.207&     0.923\\
 \fneiii& 3868.75&   1F&    19.164&     6.0&     1.131&     0.784&     0.006&     0.790&     0.816&     0.006&     0.822\\
 \fneiii& 3967.46&   1F&     5.689&     6.0&     1.147&     0.795&     0.006&     0.801&     0.828&     0.006&     0.835\\
 \fneiii&   15.55$\mu$m&    ~&   110.660&  \ldots&     1.006&     1.054&     0.203&     1.257&     1.050&     0.206&     1.255\\
 \fneiii&   36.02$\mu$m&    ~&     7.360&  \ldots&     1.275&     1.331&     0.236&     1.566&     1.326&     0.239&     1.565\\
\noalign{\vskip2pt}
   \fsii& 4068.60&   1F&     0.223&       :&     1.004&     0.699&     0.011&     0.710&     0.772&     0.012&     0.784\\
   \fsii& 6716.47&   2F&     0.957&     7.0&     0.978&     0.742&     0.026&     0.769&     0.809&     0.028&     0.837\\
   \fsii& 6730.85&   2F&     1.441&     7.0&     1.010&     0.773&     0.032&     0.805&     0.843&     0.034&     0.877\\
  \fsiii& 6312.10&   3F&     0.639&     9.0&     3.617&     1.984&     0.015&     2.000&     2.109&     0.016&     2.126\\
  \fsiii&   18.68$\mu$m&    ~&    54.820&  \ldots&     2.495&     1.974&     0.393&     2.367&     2.008&     0.395&     2.403\\
  \fsiii&   33.65$\mu$m&    ~&    30.360&  \ldots&     2.076&     1.624&     0.225&     1.849&     1.652&     0.226&     1.878\\
\noalign{\vskip2pt}
 \fcliii& 5517.71&   1F&     0.366&    14.0&     0.940&     0.711&     0.014&     0.725&     0.736&     0.014&     0.750\\
 \fcliii& 5537.88&   1F&     0.366&    14.0&     1.054&     0.806&     0.020&     0.826&     0.834&     0.021&     0.855\\
\noalign{\vskip2pt}
 \fariii& 7135.78&   1F&    15.477&     7.0&     1.048&     0.775&     0.035&     0.809&     0.796&     0.035&     0.831\\
 \fariii& 7751.10&   2F&     3.493&     7.0&     1.113&     0.822&     0.037&     0.859&     0.845&     0.038&     0.883\\
 \fariii&    8.99$\mu$m&    ~&    14.970&  \ldots&     1.657&     1.483&     0.351&     1.834&     1.490&     0.351&     1.841\\
\noalign{\vskip2pt}
\multicolumn{3}{l}{\hb\,$^{\mathrm{a}}$/10$^{-12}$ erg\,cm$^{-2}$\,s$^{-1}$}& 19.7&     13.0&      0.884&0.863     &0.255     &1.118     &0.856     &0.254     &     1.110\\
\noalign{\vskip2pt}
\multicolumn{2}{l}{$\tau$(\hi)\,$^{\mathrm{b}}$}&     &     &     &0.556     &     &     &0.581     &     &     &0.651     \\
\multicolumn{2}{l}{$\tau$(\hei)\,$^{\mathrm{c}}$}&     &     &     &0.336     &     &     &0.343     &     &     &0.422    \\
\noalign{\vskip3pt} \noalign{\hrule}\noalign{\vskip3pt}
\end{tabular}
\begin{list}{}{}
\item[$^{\mathrm{a}}$] The intrinsic H$\beta$ line flux of the entire nebula. 
\item[$^{\mathrm{b}}$] Optical depth at the \hi\ photoionization threshold (13.6\,eV). 
\item[$^{\mathrm{c}}$] Optical depth at the \hei\ photoionization threshold (24.6\,eV). 
\item[$^{\mathrm{d}}$] Recombination contribution estimated by equations (\ref{pb8:temp:nii:correction}) and (\ref{pb8:temp:oii:correction})  included in the predicted lines.
\end{list}
\end{table*}

\section{Results}
\label{pb8:sec:results}

\subsection{Comparison of the emission-line fluxes}

Table~\ref{pb8:modelresults} lists the observed and predicted nebular emission line fluxes. Column 4 presents the observed, dereddened intensities of PB\,8 from \citet{Garcia-Rojas2009}, relative to the intrinsic dereddened H$\beta$ flux, on a scale where $I$(\hb)$=100$.
The ratios of predicted over observed values from the model MC1 are presented in Column 6. Columns 7--9 present the ratios of predicted over observed values for the normal component, the metal-rich component, and  the entire nebula (normal+metal-rich) from the best-fitting model MC2. The same values obtained from the model MC3 are given in Columns 10--12. The majority of the CEL intensities predicted by model MC1 are in reasonable agreement with the observations. However, there are some large discrepancies between the prediction of model MC1 and the observations for ORLs. From the model MC2, it can be seen that the ORL discrepancy between model and observations can be explained by recombination processes of colder metal-rich inclusions embedded in the global H-rich environments.

\begin{figure*}
\begin{center}
\includegraphics[width=7.0in]{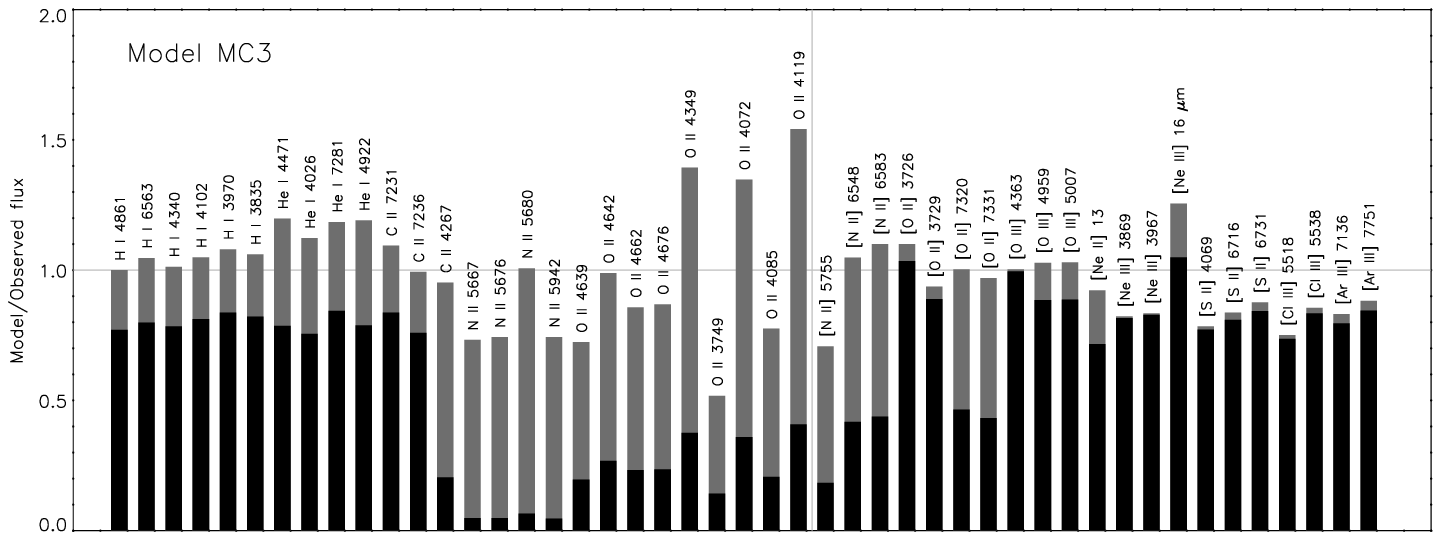}
\caption{The predicted over observed flux ratio for the bi-abundance model MC3. The relative contributions of the normal and the metal-rich components to each line flux are shown by black and grey parts, respectively.
}%
\label{pb8:mode:obs}%
\end{center}
\end{figure*}

As seen in Table~\ref{pb8:modelresults}, the $[$N\,{\sc ii}$]$ $\lambda$6584 and $[$O\,{\sc iii}$]$ $\lambda$5007 line intensities predicted by the models MC1 and MC2 are in excellent agreement with the observations. As both the models MC2 and MC3 have exactly the same density distribution and chemical abundances, we can see how dust grains introduce a $10$ percent increase in the $[$N\,{\sc ii}$]$ $\lambda$6584 line, which means that nitrogen abundance could be overestimated in some dusty nebulae. The H\,{\sc i} line intensities as well as the majority of the He\,{\sc i} line intensities are in reasonable agreement with the observations, discrepancies within 20 percent, apart from the He\,{\sc i} $\lambda$3889, $\lambda$5875 and $\lambda$7065 (around 30 percent). This could be due to high uncertainties of the recombination coefficients of the He\,{\sc i} lines below 5000\,K \citep[see ][]{Porter2012,Porter2013}. The [O\,{\sc ii}]\,$\lambda$7319 and $\lambda$7330 doublets are underestimated by around 50 percent in the model MC1. Recombination processes can largely contribute to the observed fluxes of these lines, which can be estimated by the empirical equation given by \citet{Liu2000} (see equation~\ref{pb8:temp:oii:correction}).

There are discrepancies between the predicted intensities of $[$S\,{\sc ii}$]$ and $[$S\,{\sc iii}$]$ lines and the observed values. While the intensities of the $[$S\,{\sc ii}$]$ lines are predicted to be about 10--20 percent lower than the observations, the intensity of the $[$S\,{\sc iii}$]$ $\lambda$6312 line is calculated to be almost twice more than the observed value. Adjusting the sulfur abundance cannot help reproduce $[$S\,{\sc iii}$]$ lines, so these discrepancies could be related to either the atomic data or the physical conditions. 
The predicted intensities of $[$S\,{\sc ii}$]$ lines were calculated using S$^{+}$ collision strengths from \citet{Ramsbottom1996} incorporated into the CHIANTI database (V 7.0), which is currently used in {\sc mocassin}. Recently, new S$^{+}$ collision strengths were calculated by \citet{Tayal2010},  which ignored the effect of coupling to the continuum in their calculations, so their results were estimated to be accurate to about 30 percent or better. 
We note that the emissivities of $[$S\,{\sc ii}$]$ $\lambda\lambda$6716,6731 lines calculated by the proEQUIB IDL Library \footnote{\url{https://github.com/equib/proEQUIB}}, which includes an IDL implementation of the Fortran program EQUIB \citep{Howarth1981,Howarth2016}, show that the collision strengths by \citet{Tayal2010} make them about 8 percent lower at the given physical conditions.
The predicted $[$S\,{\sc iii}$]$ line intensities are perhaps much more uncertain, as there seem to be some errors in the atomic data, as mentioned by \citet{Grieve2014}. For example, the emissivity of $[$S\,{\sc iii}$]$ 18.68$\mu$m line calculated using the collision strengths from \citet{Tayal1999} is about 40 percent higher than the calculation made with \citet{Hudson2012} or \citet{Grieve2014}. 
This issue might be related to the long-standing problem of the sulfur anomaly in PNe \citep[see reviews by][]{Henry2012}. 

The predicted intensities of the $[$Ar\,{\sc iii}$]$ $\lambda\lambda$7136,7751 lines are in agreement with the observations, discrepancies within 20 percent, however, the IR fine-structure $[$Ar\,{\sc iii}$]$ 8.99 $\mu$m line is predicted to be about 80 percent higher. We used Ar$^{2+}$ collision strengths from \citet{Galavis1995} used by the CHIANTI database (V 7.0). There is another set for Ar$^{2+}$ collision strengths \citep{MunozBurgos2009} whose predictions are significantly different and need to be examined carefully. We notice that the emissivities of $[$Ar\,{\sc iii}$]$ $\lambda\lambda$7136,7751 lines predicted by proEQUIB with the collision strengths from \citet{MunozBurgos2009} show a discrepancy of about 9 percent in comparison to those calculated with \citet{Galavis1995}, whereas there is a 30 percent difference in the $[$Ar\,{\sc iii}$]$ 8.99 $\mu$m emissivity calculated with the different atomic data.

The predicted $[$Ne\,{\sc ii}$]$ $\lambda$12.82 $\mu$m and $[$Ne\,{\sc iii}$]$ $\lambda\lambda$3869,3967 line intensities do not show high discrepancies (less than 20 percent), nevertheless, the calculated intensities of $[$Ne\,{\sc ii}$]$ $\lambda\lambda$15.55,36.02 $\mu$m lines have discrepancies about 26 and 57 percent. 
The predicted $[$Cl\,{\sc iii}$]$ $\lambda\lambda$5518,5538 lines are in agreement with the observations, discrepancies less than 25 percent.

Although the [O\,{\sc iii}]~$\lambda$4363 auroral line is perfectly matched by the model MC3 and discrepancies remain less than 10 percent in the model MC2, there is a notable discrepancy in the [N\,{\sc ii}]~$\lambda$5755 auroral line. This could be due to excitation by continuum fluorescence and/or recombination process. \citet{Bautista1999} found that the [N\,{\sc i}]~$\lambda\lambda$5198,5200 lines are efficiently affected by fluorescence excitation in many objects, while 
[O\,{\sc i}] lines were found to be sensitive to fluorescence in colder regions ($\leq 5000$\,K) or very high radiation fields. 
Nevertheless, this PN is not known to be surrounded by a photo-dissociation region (PDR) that is responsible for the fluorescence excitation. We notice that \citet{Garcia-Rojas2009} observed the brightest part of the nebula, and excluded the central star contamination and the surrounding potential PDR. Moreover, the absences of the \foi $\lambda\lambda$6300,6364 lines emitted by neutral O$^{0}$ ion and the \fni $\lambda\lambda$5198,5200 lines emitted by neutral N$^{0}$ ion in the spectrum presented by \citet{Garcia-Rojas2009} exclude any possibilities of the fluorescence contamination.
Hence, there is no strong evidence for any possible fluorescence contributions to the observed fluxes. Alternatively, the recombination contribution to [N\,{\sc ii}] auroral lines may have some implications, which can be estimated for low-density uniform nebular media \citep[see e.g.][]{Liu2000}. 

The recombination contribution to the [N\,{\sc ii}] $\lambda$5755 line and the [O\,{\sc ii}] $\lambda\lambda$7320,7330 doublet can be estimated as follows \citep{Liu2000}: 
\begin{equation}
\frac{I_{\rm R}(\lambda5755)}{I({\rm H}\beta)}=3.19 \,\, {t^{0.30}} \, \bigg( \frac{\rm N^{2+}}{\rm H^{+}} \bigg)_{\rm ORLs},
\label{pb8:temp:nii:correction}
\end{equation}
\begin{equation}
\frac{I_{\rm R}(\lambda7320+\lambda7330)}{I({\rm H}\beta)}=9.36 \,\, {t^{0.44}} \, \bigg( \frac{\rm O^{2+}}{\rm H^{+}} \bigg)_{\rm ORLs},
\label{pb8:temp:oii:correction}
\end{equation}
where $t\equiv T_{\rm e}/10^4$ is the electron temperature in $10^4$\,K from Tables \ref{pb8:tab:temperatures1} and \ref{pb8:tab:temperatures2} and $({\rm N^{2+}}/{\rm H^{+}})_{\rm ORLs}$ and $({\rm O^{2+}}/{\rm H^{+}})_{\rm ORLs}$ derived from Tables \ref{pb8:modelparameters}, \ref{pb8:tab:ionfraction1} and \ref{pb8:tab:ionfraction2}. 
The recombination contributions to the [N\,{\sc ii}]~$\lambda$5755 auroral line are estimated to be about 12 percent in the model MC1 and 48 percent in the models MC2 and MC3 (including contributions from the metal-rich inclusions).
As {\sc mocassin}  does not currently estimate the recombination contributions to the auroral lines (due to the lack of atomic data), we empirically calculated them with equations (\ref{pb8:temp:nii:correction}) and (\ref{pb8:temp:oii:correction}) and included them  in the [N\,{\sc ii}]~$\lambda$5755 line and the [O\,{\sc ii}] $\lambda\lambda$7320,7330 doublet in Tables \ref{pb8:modelresults}.
We see that the [N\,{\sc ii}]~$\lambda$5755 auroral line in the model MC3 show better agreement, but about 30 percent lower. The uncertainty of this faint line, 14 percent reported by \citet{Garcia-Rojas2009}, which could be even higher, may explain this discrepancy. Additionally, it is extremely difficult to evaluate the recombination contribution in the presence of inhomogeneous condensations. The collisional de-excitations of very dense clumps in the nebula can suppress the $\lambda\lambda$6548,6584 nebular lines, but not the auroral lines \citep{Viegas1994}, so the discrepancy between the model and the observation could be related due to unknown inhomogeneous condensations and high uncertainties in the recombination atomic data at low temperatures (i.e., below 5000\,K in a dense clump). Similarly, the recombination contribution to the [O\,{\sc ii}]~$\lambda\lambda$7320,7330 doublets are estimated to be about 15 percent in the model MC1 and 53 percent in the models MC2 and MC3. As you see, the predicted [O\,{\sc ii}]~$\lambda\lambda$7320,7330 doublets are in excellent agreement with the observations.

\begin{table}[t]
\footnotesize
\begin{center}
\caption{Mean electron temperatures (K) weighted by ionic species for the entire nebula.
For each element the first row is for MC1 and the second row is for MC2.
\label{pb8:tab:temperatures1}
}
\begin{tabular}{lccccccc}
\hline
\hline
     & \multicolumn{7}{c}{Ion}\\
\cline{2-8}
  El.~~~~ & {\sc i}   &{\sc ii}   &{\sc iii}&{\sc iv}&{\sc v} &{\sc vi}&{\sc vii}\\
\hline
\noalign{\smallskip}
H &     7746 &     7625 &   &   &   &   &   \\
  &     6647 &     6584 &   &   &   &   &   \\
He &     7746 &     7625 &     7595 &   &   &   &   \\
   &     6655 &     6584 &     6784 &   &   &   &   \\
C &     7829 &     7741 &     7621 &     7468 &     7431 &     7625 &     7625 \\
  &     6806 &     6673 &     6580 &     6549 &     6631 &     6584 &     6584 \\
N &     7834 &     7746 &     7617 &     7468 &     7431 &     7625 &     7625 \\
  &     6883 &     6772 &     6567 &     6439 &     6565 &     6584 &     6584 \\
O &     7860 &     7746 &     7613 &     7566 &     7625 &     7625 &     7625 \\
  &     7176 &     6692 &     6568 &     6618 &     6584 &     6584 &     6584 \\
Ne &     7812 &     7720 &     7601 &     7564 &     7625 &     7625 &     7625 \\
   &     6602 &     6599 &     6579 &     6764 &     6584 &     6584 &     6584 \\
S &     7855 &     7783 &     7685 &     7541 &     7411 &     7390 &     7625 \\
  &     6760 &     6676 &     6632 &     6497 &     6453 &     6572 &     6584 \\
Cl &     7836 &     7748 &     7635 &     7503 &     7467 &     7625 &     7625 \\
   &     6747 &     6658 &     6598 &     6408 &     6612 &     6584 &     6584 \\
Ar &     7846 &     7766 &     7659 &     7535 &     7490 &     7625 &     7625 \\
   &     6699 &     6606 &     6588 &     6570 &     6684 &     6584 &     6584 \\
\noalign{\smallskip}
\hline
\end{tabular}
\end{center}
\end{table}

The intensities of the ORLs predicted by the model MC2 and MC3, both bi-abundance models, can be compared to the observed values in Table~\ref{pb8:modelresults}. Figure~\ref{pb8:mode:obs} compares the predicted over observed flux ratio for the model MC3, and shows the relative contributions of the normal and the metal-rich components to each emission-line flux. The agreement between the ORL intensities predicted by the two latter models and the observations is better than those derived from the first model (MC1). The majority of the O\,{\sc ii} lines with strong intensities are in reasonable agreement with the observations, with discrepancies below 40 percent, except for $\lambda$4649.13, $\lambda$3749.48, $\lambda$4075.86, $\lambda$4132.80 and $\lambda$4153.30. The well-measured N\,{\sc ii} $\lambda$5666.64, $\lambda$5676.02 and $\lambda$5679.56 lines are in good agreement with the observations, and discrepancies are less than 30 percent. 
There are some discrepancies in some O\,{\sc ii} ORLs (e.g. $\lambda$4416.97, $\lambda$4121.46, $\lambda$4906.81) and N\,{\sc ii} ORLs (e.g. $\lambda$4601.48, $\lambda$4613.87, $\lambda$5931.78), which have weak intensities and higher uncertainties (20--30 percent). However, as seen in Figure \ref{pb8:mode:obs}, the model MC3 has significant improvements in predicting the \oii\ and \nii\ lines having intensities stronger then other ORLs. Particularly, the bi-abundance models MC2 and MC3 provide better predictions for the \oii\ ORLs from the V1 multiplet and the \nii\ ORLs from the V3 multiplet, which have the reliable atomic data. Comparing Figure~\ref{pb8:mode:obs}  with  Fig. 15 in \citet{Yuan2011} demonstrates that our bi-abundance models of PB\,8, similar to the photoionization model of NGC\,6153, better predict the observed intensities of the \nii\ and \oii\ ORLs. The models also reproduce the C~{\sc ii} ORLs with discrepancies about 10 percent, except the C\,{\sc ii} $\lambda$6578 line. The C\,{\sc ii} $\lambda$4267.2 line is stronger than the other \cii\ lines, and it is not blended with any nearby O\,{\sc ii} ORLs. The C\,{\sc ii} $\lambda$6578 line may be blended with nearby lines, so its measured line strength may be  uncertainty. 

\begin{table}[t]
\footnotesize
\begin{center}
\caption{Mean electron temperatures (K) weighted by ionic species for the nebula obtained from the photoionization model MC3. For each element the first row is for the normal component, the second row is for the H-poor component,  and the third row is for the entire nebula.
\label{pb8:tab:temperatures2}
}
\begin{tabular}{lccccccc}
\hline
\hline
     & \multicolumn{7}{c}{Ion}\\
\cline{2-8}
  El.~~~~ & {\sc i}   &{\sc ii}   &{\sc iii}&{\sc iv}&{\sc v} &{\sc vi}&{\sc vii}\\
\hline
\noalign{\smallskip}
H &     7298 &     7097 &   &   &   &   &   \\
  &     4341 &     4309 &   &   &   &   &   \\
  &     6719 &     6640 &   &   &   &   &   \\
\noalign{\smallskip}
He &     7307 &     7097 &     7054 &   &   &   &   \\
   &     4343 &     4309 &     4310 &   &   &   &   \\
   &     6726 &     6640 &     6843 &   &   &   &   \\
\noalign{\smallskip}
C &     7436 &     7295 &     7088 &     6795 &     6739 &     7098 &     7098 \\
  &     4361 &     4342 &     4307 &     4252 &     4253 &     4309 &     4309 \\
  &     6886 &     6742 &     6635 &     6491 &     6640 &     6641 &     6641 \\
\noalign{\smallskip}
N &     7460 &     7315 &     7077 &     6796 &     6738 &     7098 &     7098 \\
  &     4363 &     4344 &     4306 &     4257 &     4258 &     4309 &     4309 \\
  &     6959 &     6840 &     6622 &     6414 &     6580 &     6641 &     6641 \\
\noalign{\smallskip}
O &     7507 &     7319 &     7064 &     6989 &     7098 &     7098 &     7098 \\
  &     4364 &     4346 &     4302 &     4302 &     4309 &     4309 &     4309 \\
  &     7250 &     6762 &     6621 &     6671 &     6641 &     6641 &     6641 \\
\noalign{\smallskip}
Ne &     7414 &     7260 &     7042 &     6988 &     7098 &     7098 &     7098 \\
   &     4361 &     4340 &     4294 &     4295 &     4309 &     4309 &     4309 \\
   &     6690 &     6671 &     6630 &     6819 &     6641 &     6641 &     6641 \\
\noalign{\smallskip}
S &     7466 &     7349 &     7186 &     6937 &     6713 &     6678 &     7098 \\
  &     4365 &     4349 &     4324 &     4276 &     4243 &     4243 &     4309 \\
  &     6848 &     6751 &     6694 &     6541 &     6435 &     6597 &     6641 \\
\noalign{\smallskip}
Cl &     7444 &     7302 &     7113 &     6875 &     6720 &     6721 &     7098 \\
   &     4361 &     4343 &     4312 &     4261 &     4241 &     4245 &     4309 \\
   &     6832 &     6731 &     6656 &     6444 &     5640 &     6241 &     6641 \\
\noalign{\smallskip}
Ar &     7468 &     7336 &     7148 &     6929 &     6861 &     7098 &     7098 \\
   &     4366 &     4349 &     4317 &     4267 &     4267 &     4309 &     4309 \\
   &     6783 &     6680 &     6647 &     6618 &     6737 &     6641 &     6641 \\
\noalign{\smallskip}
\hline
\end{tabular}
\end{center}
\end{table}

\subsection{Thermal structure}
\label{pb8:sec:results:temp}

Table~\ref{pb8:tab:temperatures1} lists the mean electron temperatures of the entire nebula in the models MC1 and MC2 weighted by ionic species, from the neutral ({\sc i}) to the highly ionized ions ({\sc vii}). The definition for the weighted-mean temperatures was given in \citet{Ercolano2003a}. 
The value of $T_{\rm e}($N\,{\sc ii}$)=7746$\,K predicted by the model MC1 is about 1150\,K lower than the value of $T_{\rm e}($N\,{\sc ii}$)=8900 \pm 500$\,K empirically derived from CELs by \citet{Garcia-Rojas2009}. This could be due to recombination contributions to the auroral line. The recombination contribution estimated \citep[using Eq. 1 in][]{Liu2000} about 12 percent in the model MC1 is much lower to explain the measured intensity of the [N\,{\sc ii}]~$\lambda$5755 line. However, the model MC2 predicts the [N\,{\sc ii}]~$\lambda$5755 line to be 40 percent higher than the value derived from the model MC1 (see recombination contribution estimated by Eq.~\ref{pb8:temp:nii:correction} in Table~\ref{pb8:modelresults}). We also notice that the [N\,{\sc ii}] temperature is roughly equal to the [O\,{\sc iii}] temperature in low excitation PNe  \citep{Kingsburgh1994}, so the empirical value of the [N\,{\sc ii}] electron temperature derived by \citet{Garcia-Rojas2009} is difficult to be explained. The temperature $T_{\rm e}$(O\,{\sc ii}$)=7746$\,K predicted by the model MC1 is about 670\,k higher than $T_{\rm e}($O\,{\sc ii}$)=7050 \pm 400$\,K empirically derived by \citet{Garcia-Rojas2009}, while $T_{\rm e}$(O\,{\sc ii}$)=6692$\,K predicted by the model MC2 is about 360\,k lower than the empirical value.
Moreover, the temperature of [O\,{\sc iii}] calculated from the model MC1, $T_{\rm e}($O\,{\sc iii}$)=7613$\,K, is about 710\,K higher the empirical result of $T_{\rm e}($O\,{\sc iii}$)=6900 \pm 150$\,K, whereas $T_{\rm e}($O\,{\sc iii}$)=6568$\,K predicted by the model MC2 is about 330\,K lower the empirical value.

Table~\ref{pb8:tab:temperatures2} presents the electron temperatures of the different components of the model MC3 weighted by ionic abundances, as well as the mean temperatures of the entire nebula. The first entries for each element are for the normal abundance plasma, the second entries are for the metal-rich inclusion, and the third entries are for the entire nebula (including both the normal and the H-poor components). It can be seen that the temperatures weighted by ionic abundances in the two different components of the nebula are very different. The electron temperatures separately weighted by the ionic species of the metal-rich inclusions were much lower than those from the normal part. The temperature of $T_{\rm e}($N\,{\sc ii}$)=7315$\,K predicted by the normal component of the model MC3 is about 1590\,K lower than the value empirically derived. 
However, $T_{\rm e}$(O\,{\sc ii}$)=7319$\,K and $T_{\rm e}$(O\,{\sc iii}$)=7064$\,K obtained by the normal component of the model MC3 are in reasonable agreement with the values empirically derived by \citet{Garcia-Rojas2009}. We see that $T_{\rm e}($He\,{\sc i}$)=4309$\,K weighted by the metal-rich component of the model MC3 is lower than the empirical value, $T_{\rm e}($He\,{\sc i}$)=6250 \pm 150$\,K \citep{Garcia-Rojas2009}, whereas $T_{\rm e}($He\,{\sc i}$)=6640$\,K weighted by the whole nebula is about 390\,K higher the empirical value
Moreover, $T_{\rm e}($H\,{\sc i}$)=4309$\,K weighted by the metal-rich component and $T_{\rm e}($H\,{\sc i}$)=6640$\,K weighted by the whole nebula are reasonably in the range of $T_{\rm e}($H\,{\sc i}$)=5100^{+1300}_{-900}$\,K empirically derived from the Balmer Jump to H11 flux ratio \citep{Garcia-Rojas2009}.  We take no account for the interaction between the two components, namely normal and metal-rich, which could also lead to a temperature variation. Note that the radiative transfer in a neutral region is not currently supported by MOCASSIN, so the code only estimates temperatures that likely correspond to a potential narrow transition region between ionized and neutral regions for a radiation-bounded object. Neutral elements in this PN are negligible (see Table~\ref{pb8:tab:ionfraction1}), so the temperatures of the neutral species listed in Table~\ref{pb8:tab:temperatures2} do not have any significant physical meaning. We also see that the two components have a local thermal pressure ratio of $P($metal-rich$)/P($normal$) \sim 1.1$, which means each metal-rich cell is in pressure equilibrium with its surrounding normal gas. The higher thermal pressure forces the dense, metal-rich knots to expand and reduce their density and temperature during the evolution phase of the nebula.

\subsection{Fractional ionic abundances}
\label{pb8:sec:results:ionic}

\begin{table*}
\footnotesize
\begin{center}
\caption{Fractional ionic abundances obtained from the photoionization models.
For each element the first row is for MC1, the second row is for MC2 and the third row is for MC3.
\label{pb8:tab:ionfraction1}
}
\begin{tabular}{llcccccc}
\multicolumn{8}{c}{}\\
\hline
\hline
   & \multicolumn{7}{c}{Ion}\\
\cline{2-8}
Element & {\sc i}   &{\sc ii}   &{\sc iii}&{\sc iv}&{\sc v} &{\sc vi}&{\sc vii}\\
\hline
\noalign{\smallskip}
H  & 6.88($-4$) & 9.99($-1$) &   &   &   &   &   \\
   & 8.23($-4$) & 9.99($-1$) &   &   &   &   &   \\
   & 8.59($-4$) & 9.99($-1$) &   &   &   &   &   \\
\noalign{\smallskip}
He & 1.91($-3$) & 9.98($-1$) & 1.62($-12$) &   &   &   &   \\
   & 2.39($-3$) & 9.98($-1$) & 1.38($-12$) &   &   &   &   \\
   & 2.47($-3$) & 9.98($-1$) & 1.39($-12$) &   &   &   &   \\
\noalign{\smallskip}
C  & 1.19($-5$) & 4.15($-2$) & 9.56($-1$) & 2.39($-3$) & 2.70($-16$) & 1.00($-20$) & 1.00($-20$) \\
   & 1.63($-5$) & 5.03($-2$) & 9.48($-1$) & 1.86($-3$) & 1.91($-16$) & 1.00($-20$) & 1.00($-20$) \\
   & 1.75($-5$) & 5.18($-2$) & 9.46($-1$) & 1.71($-3$) & 1.75($-16$) & 1.00($-20$) & 1.00($-20$) \\
\noalign{\smallskip}
N  & 1.60($-5$) & 6.82($-2$) & 9.28($-1$) & 3.40($-3$) & 6.86($-16$) & 1.00($-20$) & 1.00($-20$) \\
   & 2.51($-5$) & 8.50($-2$) & 9.12($-1$) & 2.92($-3$) & 5.41($-16$) & 1.00($-20$) & 1.00($-20$) \\
   & 2.69($-5$) & 8.74($-2$) & 9.10($-1$) & 2.80($-3$) & 5.21($-16$) & 1.00($-20$) & 1.00($-20$) \\
\noalign{\smallskip}
O  & 6.40($-5$) & 9.40($-2$) & 9.06($-1$) & 2.00($-13$) & 1.00($-20$) & 1.00($-20$) & 1.00($-20$) \\
   & 1.06($-4$) & 1.31($-1$) & 8.68($-1$) & 1.82($-13$) & 1.00($-20$) & 1.00($-20$) & 1.00($-20$) \\
   & 1.19($-4$) & 1.36($-1$) & 8.64($-1$) & 1.81($-13$) & 1.00($-20$) & 1.00($-20$) & 1.00($-20$) \\
\noalign{\smallskip}
Ne & 1.06($-4$) & 2.02($-1$) & 7.97($-1$) & 7.74($-14$) & 1.00($-20$) & 1.00($-20$) & 1.00($-20$) \\
   & 1.78($-4$) & 2.61($-1$) & 7.39($-1$) & 6.04($-14$) & 1.00($-20$) & 1.00($-20$) & 1.00($-20$) \\
   & 1.87($-4$) & 2.65($-1$) & 7.35($-1$) & 6.03($-14$) & 1.00($-20$) & 1.00($-20$) & 1.00($-20$) \\
\noalign{\smallskip}
S  & 3.01($-7$) & 4.72($-3$) & 5.81($-1$) & 4.13($-1$) & 1.90($-3$) & 8.26($-16$) & 1.00($-20$) \\
   & 4.46($-7$) & 6.14($-3$) & 6.36($-1$) & 3.56($-1$) & 1.46($-3$) & 5.63($-16$) & 1.00($-20$) \\
   & 4.87($-7$) & 6.46($-3$) & 6.42($-1$) & 3.50($-1$) & 1.42($-3$) & 5.48($-16$) & 1.00($-20$) \\
\noalign{\smallskip}
Cl  & 2.68($-6$) & 1.85($-2$) & 8.91($-1$) & 9.02($-2$) & 9.99($-15$) & 1.00($-20$) & 1.00($-20$) \\
    & 3.79($-6$) & 2.22($-2$) & 8.97($-1$) & 8.09($-2$) & 7.66($-15$) & 1.00($-20$) & 1.00($-20$) \\
    & 4.12($-6$) & 2.30($-2$) & 8.96($-1$) & 8.09($-2$) & 5.00($-7$) & 1.47($-19$) & 1.00($-20$) \\
\noalign{\smallskip}
Ar & 4.47($-7$) & 4.11($-3$) & 7.21($-1$) & 2.75($-1$) & 1.56($-13$) & 1.00($-20$) & 1.00($-20$) \\
   & 7.85($-7$) & 5.95($-3$) & 7.72($-1$) & 2.22($-1$) & 1.11($-13$) & 1.00($-20$) & 1.00($-20$) \\
   & 8.43($-7$) & 6.12($-3$) & 7.72($-1$) & 2.22($-1$) & 1.12($-13$) & 1.00($-20$) & 1.00($-20$) \\
\noalign{\smallskip}
\hline
\end{tabular}
\end{center}
\end{table*}

\begin{table*}
\footnotesize
\begin{center}
\caption{Fractional ionic abundances obtained from the photoionization model MC2.
For each element the first row is for the normal component and the second row is for the H-poor component.
\label{pb8:tab:ionfraction2}
}
\begin{tabular}{llcccccc}
\multicolumn{8}{c}{}\\
\hline
\hline
   & \multicolumn{7}{c}{Ion}\\
\cline{2-8}
Element & {\sc i}   &{\sc ii}   &{\sc iii}&{\sc iv}&{\sc v} &{\sc vi}&{\sc vii}\\
\hline
\noalign{\smallskip}
H  & 7.86($-4$) & 9.99($-1$) &   &   &   &   &   \\
   & 1.01($-3$) & 9.99($-1$) &   &   &   &   &   \\
\noalign{\smallskip}
He & 2.28($-3$) & 9.98($-1$) & 1.53($-12$) &   &   &   &   \\
   & 2.93($-3$) & 9.97($-1$) & 6.49($-13$) &   &   &   &   \\
\noalign{\smallskip}
C  & 1.59($-5$) & 4.86($-2$) & 9.49($-1$) & 2.05($-3$) & 2.22($-16$) & 1.00($-20$) & 1.00($-20$) \\
   & 1.87($-5$) & 5.89($-2$) & 9.40($-1$) & 9.20($-4$) & 3.12($-17$) & 1.00($-20$) & 1.00($-20$) \\
\noalign{\smallskip}
N  & 2.49($-5$) & 8.49($-2$) & 9.12($-1$) & 3.06($-3$) & 6.13($-16$) & 1.00($-20$) & 1.00($-20$) \\
   & 2.60($-5$) & 8.55($-2$) & 9.12($-1$) & 2.20($-3$) & 1.73($-16$) & 1.00($-20$) & 1.00($-20$) \\
\noalign{\smallskip}
O  & 1.16($-4$) & 1.27($-1$) & 8.73($-1$) & 1.92($-13$) & 1.00($-20$) & 1.00($-20$) & 1.00($-20$) \\
   & 5.74($-5$) & 1.55($-1$) & 8.45($-1$) & 1.31($-13$) & 1.00($-20$) & 1.00($-20$) & 1.00($-20$) \\
\noalign{\smallskip}
Ne & 1.60($-4$) & 2.47($-1$) & 7.53($-1$) & 6.77($-14$) & 1.00($-20$) & 1.00($-20$) & 1.00($-20$) \\
   & 2.72($-4$) & 3.32($-1$) & 6.68($-1$) & 2.29($-14$) & 1.00($-20$) & 1.00($-20$) & 1.00($-20$) \\
\noalign{\smallskip}
S  & 4.20($-7$) & 5.83($-3$) & 6.29($-1$) & 3.64($-1$) & 1.59($-3$) & 6.56($-16$) & 1.00($-20$) \\
   & 5.74($-7$) & 7.74($-3$) & 6.75($-1$) & 3.17($-1$) & 7.82($-4$) & 8.85($-17$) & 1.00($-20$) \\
\noalign{\smallskip}
Cl  & 3.59($-6$) & 2.13($-2$) & 8.97($-1$) & 8.13($-2$) & 8.59($-15$) & 1.00($-20$) & 1.00($-20$) \\
    & 4.84($-6$) & 2.69($-2$) & 8.94($-1$) & 7.89($-2$) & 2.87($-15$) & 1.00($-20$) & 1.00($-20$) \\
\noalign{\smallskip}
Ar & 7.21($-7$) & 5.50($-3$) & 7.60($-1$) & 2.35($-1$) & 1.26($-13$) & 1.00($-20$) & 1.00($-20$) \\
   & 1.11($-6$) & 8.25($-3$) & 8.37($-1$) & 1.55($-1$) & 3.19($-14$) & 1.00($-20$) & 1.00($-20$) \\
\noalign{\smallskip}
\hline
\end{tabular}
\end{center}
\end{table*}

Table~\ref{pb8:tab:ionfraction1} lists the volume-averaged fractional ionic abundances from the neutral ({\sc i}) to the highly ionized ions ({\sc vii}) calculated from the three models, where, the first entries for each element are for the chemically homogeneous model MC1, the second entries are for the bi-abundance model MC2, and the third entries are the dusty bi-abundance model MC3. The definition for the volume-averaged fractional ionic abundances was given in \citet{Ercolano2003a}. 
We see that both hydrogen and helium are fully singly-ionized, i.e. neutrals are almost zero percent in the three models. It can be seen that the ionization structure in MC2 is in reasonable agreement with MC1. The elemental oxygen largely exists as O$^{2+}$ with 91 percent and then O$^{+}$ with 9 percent in the model MC1, whereas O$^{2+}$ is about 87 percent and then O$^{+}$ is about 13 percent in the model MC2. Moreover, the elemental nitrogen largely exists as N$^{2+}$ with 93 percent and then N$^{+}$ with 7 percent in the model MC1, whereas N$^{2+}$ is about 91 percent and then N$^{+}$ is about 9 percent in the model MC2. The O$^{+}$/O ratio is about 1.4--1.6 times higher than the N$^{+}$/N ratio, which is in disagreement with the general assumption of N/N$^+$=O/O$^+$ in the ionization correction factor (\textit{icf}) method by \citet{Kingsburgh1994}, introducing errors to empirically derived elemental abundances. Our (N$^{+}$/N)/(O$^{+}$/O) ratio is in agreement with the value of 0.6--0.7 predicted by the photoionization model of NGC 7009 implemented using {\sc mocassin} \citep{Gonccalves2006}. 
While the assumption N/N$^+$=O/O$^+$ overestimates the N/H elemental abundance, the new \textit{icf}(N/O) calculated using 1-D photoionization modeling provides a better agreement \citep{Delgado-Inglada2014}. 
Moreover, the O$^{2+}$/O ratio is about 1.1--1.2 higher than the Ne$^{2+}$/Ne ratio, in reasonable agreement with the assumption for the $icf$(Ne). The ionic fraction of S, Cl and Ar predicted by MC2 are approximately about the values calculated by MC1.  The small discrepancies in fractional ionic abundances between MC1 and MC2 can be explained by a small fraction of the metal-rich structures included in MC2.

The volume-averaged fractional ionic abundances calculated from the model MC2 are listed in Table~\ref{pb8:tab:ionfraction2}, the upper entries for each element in the table are for the normal component and the lower entries are for the metal-rich component of the nebula. It can be seen that the model MC2 predict different ionic fractions of O$^{+}$ for the two components of the nebula, whereas roughly the same value for N$^{+}$.
The O$^{+}$/O ratios in the metal-rich component are about 20 percent higher than those in the normal component. This means that that the ionization correction factors (\textit{icf}) from CELs are not entirely accurate for deriving the elemental abundances from ORLs as adopted by some authors \citep[see e.g.][]{Wang2007}.

\section{Conclusions}
\label{pb8:sec:conclusions}

Three photoionization models have been constructed for the planetary nebula PB\,8, a chemically homogeneous model, a bi-abundance model and a dusty bi-abundance model. Our intention was to construct a model that well reproduce the observed emission-lines and thermal structure determined from the plasma diagnostics. A powerlaw radial density profile was adopted for the spherical nebula distribution based on the radiation-hydrodynamics simulations. The density model parameters were adjusted to reproduce the total H$\beta$ intrinsic line flux of the nebula, and the mean electron density empirically derived from the CELs  \citep[][]{Garcia-Rojas2009}. We have used the non-LTE model atmosphere derived by \citet{Todt2010} for temperature $T_{\rm eff}=52$\,kK and luminosity $L_{\rm \star}=6000$L$_{\odot}$. This ionizing source well reproduced the nebular observed H$\beta$ absolute flux, as well as the $[$O\,{\sc iii}$]$ $\lambda$5007 line flux, at the distance of 4.9~kpc.

Our initial model reproduces the majority of CELs and the thermal structure, but large discrepancies exist in the observed ORLs from heavy element ions. It is found that a chemically homogeneous model cannot consistently explain the ORLs observed in the nebular spectrum. We therefore intended to address the cause of the heavily underestimated ORLs. Following the hypothesis of the bi-abundance model by \citet{Liu2000}, a small fraction of metal-rich inclusions was introduced into the second model. The heavy element ORLs are mostly emitted from the metal-rich structures embedded in the dominant diffuse warm plasma of normal abundances.
The agreement between the ORL intensities predicted by the model MC2 and the observations is better than the first model (MC1). The metal-rich inclusions occupying 5.6 percent of the total volume of the nebula, and are about 1.7 times cooler and denser than the normal composition nebula. The O/H and N/H abundance ratios in the metal-rich inclusions are $\sim$\,1.0 and 1.7\,dex larger than the diffuse warm nebula, respectively. The mean electron temperatures predicted by MC2 are lower than those from MC1, which is because of the cooling effects of the metal-rich inclusions. The results indicate that a bi-abundance model can naturally explain the heavily underestimated ORLs in the chemically homogeneous model. Therefore, the metal-rich inclusions may solve the problem of ORL/CEL abundance discrepancies.
However, the model MC2 cannot explain the thermal SED of the nebula observed with the \textit{Spitzer} spectrograph. In our final model, we have incorporated a dual dust chemistry consisting of two different grains, amorphous carbon and crystalline silicate, and discrete grain radii. It is found that a dust-to-gas ratio of 0.01 by mass for the whole nebula can roughly reproduce the observed IR continuum. 

The PN PB\,8 shows moderate ADFs \citep[$\sim$\,1.9--2.6;][]{Garcia-Rojas2009}, which are typical of most PNe \citep[see e.g.][]{Liu2006a}. 
Previously, the bi-abundance model were only examined in two PNe with extremely large ADFs: Abell~30  \citep{Ercolano2003b} and NGC~6153 \citep{Yuan2011}. In Abell~30, \citet{Ercolano2003b} used a metal-rich core whose density is about six times larger than the surrounding nebula. In NGC~6153, \citet{Yuan2011} used super-metal-rich knots distributed in the inner region of the nebula. In the present study, we adopted a bi-abundance model whose metal-rich knots are homogeneously distributed inside the diffuse warm nebula, and are associated with a gas-filling factors of 0.056. 
To reproduce the spectrum of PB\,8, it is not require to have extremely dense and super-metal-rich knots, since the ORLs do not correspond to very cold temperatures and extremely large ADFs such as Abell~30 and NGC~6153. 
We should mention that the stellar temperatures of Abell~30 ($T_{\rm eff}=$\,130\,kK) and NGC~6153 ($T_{\rm eff}=$\,90\,kK) are higher than that of PB\,8 ($T_{\rm eff}=$\,52\,kK), so the central star of PB\,8 is likely in an early stage of its stellar evolution towards a white dwarf in comparison with Abell~30 and NGC~6153.
Accordingly, the planetary nebula PB\,8 could be younger and less evolved than the PNe Abell~30 and NGC~6153. 
More recently, it has been found that PNe with ADFs larger than 10 mostly contain close-binary central stars \citep{Corradi2015,Jones2016,Wesson2017}. Currently, there is no evidence for a close-binary central star in PB\,8.

Our analysis showed that the bi-abundance hypothesis, which was previously tested in a few PNe with very large abundance discrepancies, could also be used to explain moderate discrepancies between ORL and CEL abundances in most of typical PNe \citep[ADFs\,$\sim$\,1.6--3.2;][]{Liu2006a}. It is unclear whether there is any link between the supposed metal-rich inclusions within the nebula and hydrogen-deficient stars. It has been suggested that a (very-) late thermal pulse is responsible for the formation of H-deficient central stars of planetary nebulae  \citep[see e.g.][]{Blocker2001,Herwig2001,Werner2001,Werner2006}. Thermal pulses normally occur during the AGB phase, when the helium-burning shell becomes thermally unstable. The (very-) late thermal pulse occurs when the star moves from the AGB phase towards the white dwarf. It returns the star to the AGB phase and makes a H-deficient stellar surface, so called \textit{born-again} scenario. However, the metal-rich component with C/O $< 1$ predicted by our photoionization models is in disagreement with the products of a born-again event \citep{Herwig2001,Althaus2005,Werner2006}. 

It is also possible that the metal-rich inclusions were introduced by other mechanism such as the evaporation and destruction of planets by stars \citep{Liu2003}. Recently, \citet{Nicholls2012,Nicholls2013} proposed that a non-Maxwellian distribution of electron energies could explain the abundance discrepancy. However, \citet{Zhang2014} found that both the scenarios, bi-abundance models and non-Maxwellian distributed electrons, are adequately consistent with observations of four PNe with very large ADFs. It is unclear whether chemically inhomogeneous plasmas introduce non-Maxwell-Boltzmann equilibrium electrons to the nebula. Alternatively, the binarity characteristics such as the orbital separation and companion masses may have a leading role in forming different abundance discrepancies in those PNe with binary central stars \citep[see e.g.][]{Herwig2001,Althaus2005}. Further studies are necessary to trace the origin of possible metal-rich knots within the nebula and the cause of various abundance discrepancies in planetary nebulae.

\begin{acknowledgements}
AD is supported by a Macquarie University Research Excellence Scholarship (MQRES) and a Sigma Xi Grants-in-Aid of Research (GIAR). 
AD thanks the anonymous referee for helpful suggestion and comments that improved the paper, Helge Todt for providing the PoWR models for expanding atmospheres and valuable comments,  Jorge~Garc{\'\i}a-Rojas and Miriam Pe{\~n}a for providing the Magellan Telescope data and helpful comments, 
Ralf Jacob for sharing results from radiation-hydrodynamics simulations and valuable suggestions, Christophe Morisset for illuminating discussions and corrections, Quentin A. Parker for providing the 2010 ANU 2.3-m data, 
David J. Frew for helping with the ANU observing proposal, 
Kyle DePew for carrying out the 2010 ANU observing run, 
and Roger~Wesson for sharing the program EQUIB (originally written by I.~D.~Howarth and S.~Adams at UCL). 
The computations in this paper were run on the NCI National Facility in Canberra, Australia, which is supported by the Australian Commonwealth Government, the swinSTAR supercomputer at Swinburne University of Technology, and the Odyssey cluster supported by the FAS Division of Science, Research Computing Group at Harvard University.
\end{acknowledgements}

\footnotesize

\end{document}